\documentclass[prd,showkeys,floatfix,twocolumn,amsmath,amssymb,floatfix]{revtex4}
\usepackage{graphicx}
\usepackage{epstopdf}
\usepackage{multirow}
\usepackage{subfigure}
\usepackage[colorlinks,citecolor=blue,anchorcolor=red,menucolor=red,linkcolor=red,filecolor=red,runcolor=red,urlcolor=blue,frenchlinks=red]{hyperref}
\newcommand{\feynp}[1]{#1\kern-0.45em/}
\allowdisplaybreaks[3]
\begin{document}

\title{$P$-wave single charmed baryons of the $SU(3)$ flavor $\bf\bar3_F$}

\author{Xuan Luo}
\author{Yi-Jie Wang}
\author{Hua-Xing Chen}
\email{hxchen@seu.edu.cn}
\affiliation{School of Physics, Southeast University, Nanjing 210094, China}
\begin{abstract}
We study the $P$-wave single charmed baryons of the $SU(3)$ flavor $\bf\bar3_F$ within the framework of heavy quark effective theory. We systematically calculate their strong and radiative decay properties using the light-cone sum rule method. Besides the $\Lambda_c(2595)$, $\Lambda_c(2625)$, $\Xi_c(2790)$, and $\Xi_c(2815)$, our results suggest the existence of two additional $\Lambda_c$ baryons and two additional $\Xi_c$ baryons. Their masses, mass splittings within the same multiplets, and decay properties are summarized in Table~\ref{tab:result} for future experimental searches.
\end{abstract}
\pagenumbering{arabic}
\maketitle

\section{Introduction}
The single charmed baryon, interpreted as a three-quark bound state consisting of one $charm$ quark and two light $up/down/strange$ quarks within the framework of the conventional quark model~\cite{Gell-Mann:1964ewy,Zweig:1964ruk,Ebert:1996ec,Gerasyuta:1999pc}, serves as an ideal platform for investigating the fine structure of hadron spectra~\cite{Chen:2016spr,Copley:1979wj,Karliner:2008sv}. Over the past several decades, substantial progress has been made in their experimental searches. For example, researchers have successfully identified all the ground-state single charmed baryons and have observed many excited ones:
\begin{itemize}

\item In 1993 the ARGUS Collaboration observed the $\Lambda_c(2625)$ baryon in the $\Lambda_c \pi \pi $ channel~\cite{ARGUS:1993vtm}, which was later confirmed by CLEO~\cite{CLEO:1994oxm}. In addition, this CLEO experiment observed another excited charmed baryon $\Lambda_c(2595)$ in the same channel, which was subsequently confirmed by Fermilab~\cite{E687:1995srl} and ARGUS~\cite{ARGUS:1997snv}. Their masses and widths were measured to be~\cite{PDG2024}: 
\begin{eqnarray}
\Lambda_c(2595)^+ : \rm M &=& 2592.25\pm 0.28{\rm~MeV} \, ,
 \nonumber
\\       \Gamma &=&2.59\pm 0.30 \pm 0.47~\rm MeV\, ;
\\ \Lambda_c(2625)^+ :\rm M &=& 2628.00\pm 0.15{\rm~MeV} \, ,
\nonumber
\\       \Gamma &<& 0.52{\rm~MeV}~{\rm at}~90\%~{\rm CL} \, .
\end{eqnarray}

\item In 1999 the CLEO Collaboration observed the $\Xi_c(2790)$ in the $\Xi_c^{\prime}\pi$ channel~\cite{CLEO:1999msf}, which was later confirmed by Belle~\cite{Belle:2008yxs}. In 2000 the CLEO Collaboration further observed the $\Xi_c(2815)$ in the $\Xi_c\pi\pi$ channel~\cite{CLEO:2000ibb}, which was also confirmed by Belle~\cite{Belle:2008yxs}. Their masses and widths were measured to be~\cite{PDG2024}: 
\begin{eqnarray}
\Xi_c(2790)^0 :\rm M &=& 2793.9\pm 0.5 {\rm~MeV} \, ,
 \nonumber
\\       \Gamma &=&10.0\pm 0.7\pm 0.8~\rm MeV\, ;
\\
\Xi_c(2790)^+ :\rm M &=& 2791.9\pm 0.5 {\rm~MeV} \, ,
 \nonumber
\\       \Gamma &=&8.9\pm 0.6\pm 0.8~\rm MeV\, ;
\\
\Xi_c(2815)^0 :\rm M &=& 2819.79\pm 0.30{\rm~MeV} \, ,
\nonumber
\\       \Gamma &=&2.54\pm 0.18 \pm 0.17{\rm~MeV} \,;
\\
\Xi_c(2815)^+ :\rm M &=& 2816.51\pm 0.25{\rm~MeV} \, ,
\nonumber
\\       \Gamma &=&2.43\pm 0.20 \pm 0.17{\rm~MeV} \, .
\end{eqnarray}

\end{itemize}
These states are good candidates for the $P$-wave single charmed baryons of the $SU(3)$ flavor $\mathbf{\bar3}_F$, whose detailed discussions can be found in the reviews~\cite{Crede:2013kia,Chen:2016spr,Cheng:2021qpd,Chen:2022asf}. To understand their internal structure, various phenomenological methods and models have been applied, including various quark models~\cite{Capstick:1986bm,Korner:1994nh,Ivanov:1998qe,Tawfiq:1998nk,Tawfiq:1999vz,Hussain:1999sp,Albertus:2005zy,Ebert:2007nw,Garcilazo:2007eh,Roberts:2007ni,Zhong:2007gp,Valcarce:2008dr,Ebert:2011kk,Ortega:2012cx,Yoshida:2015tia,Nagahiro:2016nsx,Lu:2020ivo,Chen:2021eyk}, Lattice QCD~\cite{Bowler:1996ws,Burch:2008qx,Brown:2014ena,Padmanath:2013bla,Padmanath:2017lng,Bahtiyar:2020uuj}, QCD sum rules~\cite{Yang:2020zjl,Chen:2015kpa,Bagan:1991sg,Neubert:1991sp,Broadhurst:1991fc,Huang:1994zj,Dai:1996yw,Groote:1996em,Colangelo:1998ga,Huang:2000tn,Mao:2015gya, Zhu:2000py,Lee:2000tb,Wang:2003zp,Duraes:2007te,Liu:2007fg,Zhang:2008pm,Wang:2010it,Zhou:2014ytp,Zhou:2015ywa,Wang:2017zjw,Aliev:2018ube,Xu:2020ofp,Chen:2017sci,Agaev:2017lip,Agaev:2017nn}, the $^3P_0$ model~\cite{Chen:2007xf,Chen:2017gnu,Zhao:2017fov,Ye:2017yvl}, chiral perturbation theory~\cite{Lu:2014ina,Cheng:2015naa,Cheng:2006dk,Pirjol:1997nh,Chiladze:1997ev,Blechman:2003mq,Huang:1995ke}, MIT bag model~\cite{Hwang:2006df}, and Bethe-Salpeter formalism~\cite{Guo:2007qu}, etc.

We have studied the $P$-wave charmed baryons of the $SU(3)$ flavor $\mathbf{\bar3}_F$ within the framework of heavy quark effective theory (HQET)~\cite{Eichten:1989zv,Grinstein:1990mj,Falk:1990yz}:
\begin{itemize}

\item In Ref.~\cite{Chen:2015kpa} we applied the QCD sum rule method~\cite{Gimenez:2005nt,Shifman:1978bx,Shifman:1978by,Reinders:1984sr,Narison:2002woh,Nielsen:2009uh,Gubler:2018ctz} to systematically study their mass spectrum.

\item In Ref.~\cite{Chen:2017sci} we applied the light-cone sum rule method~\cite{Braun:1988qv,Chernyak:1990ag,Ball:1998je,Ball:2006wn,Ball:2004rg,Ball:1998kk,Ball:1998sk,Ball:1998ff,Ball:2007rt,Ball:2007zt,Ball:2002ps} to study their $S$-wave decays into ground-state charmed baryons with light pseudoscalar/vector mesons.

\end{itemize}
In order to conduct a comprehensive investigation, we still need to study:
\begin{itemize}

\item their $D$-wave decays into ground-state charmed baryons with light pseudoscalar mesons.

\item their radiative transitions into ground-state charmed baryons with photons.

\end{itemize}
These calculations will be carried out in the present study. Note that the radiative transitions $\Xi_c(2815)/\Xi_c(2790) \to \Xi_c \gamma$ have been observed in the Belle experiment~\cite{Belle:2020ozq}, and more radiative transitions are expected to be observed in future Belle-II and LHC experiments. On the theoretical side, the radiative decays of ground-state charmed baryons have been extensively studied in Refs.~\cite{Zhu:1998ih,Cheng:1992xi,Wang:2009ic,Wang:2009cd,Jiang:2015xqa,Aliev:2014bma,Aliev:2009jt,Aliev:2016xvq,Aliev:2011bm,Chow:1995nw,Bahtiyar:2016dom,Bahtiyar:2015sga,Ivanov:1998wj,Savage:1994wa,Banuls:1999br,Cho:1994vg,Tawfiq:1999cf,Ivanov:1999bk,Wang:2017kfr,Gamermann:2010ga,Ortiz-Pacheco:2023kjn,Aliev:2019lfs,Peng:2024pyl,Garcia-Tecocoatzi:2025fxp}, whereas the radiative decays of excited ones have been studied to a much lesser extent~\cite{Tawfiq:1999cf,Ivanov:1999bk,Wang:2017kfr,Gamermann:2010ga,Ortiz-Pacheco:2023kjn,Aliev:2019lfs,Peng:2024pyl,Garcia-Tecocoatzi:2025fxp}.

\begin{figure*}[hbtp]
\begin{center}
\includegraphics[width=0.95\textwidth]{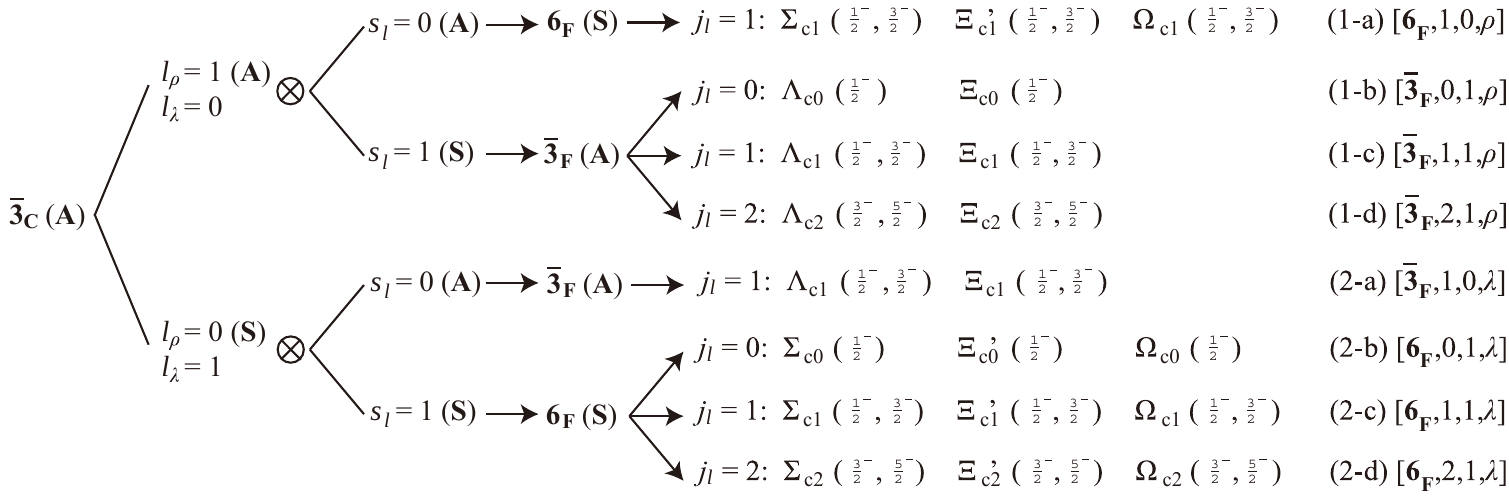}
\end{center}
\caption{Categorization of $P$-wave single charmed baryons.}
\label{fig:pwave}
\end{figure*}

In Refs.~\cite{Wang:2024rai,Luo:2024jov} we have systematically studied the strong and radiative decay properties of the $P$-wave bottom baryons belonging to the $SU(3)$ flavor $\mathbf{\bar3}_F$ representation. In the present study we just need to replace the bottom quark with the charm quark and redo the calculations. Additionally, we shall investigate the mixing effect between different HQET multiplets. This effect was not considered in our previous studies of $P$-wave bottom baryons~\cite{Wang:2024rai,Yang:2020zrh}, while it was taken into account in our study of $P$-wave charmed baryons belonging to the $SU(3)$ flavor $\mathbf{6}_F$ representation~\cite{Yang:2021lce}, {\it i.e.},
\begin{itemize}

\item $P$-wave single bottom baryons of the $SU(3)$ flavor $\mathbf{6}_F$ was systematically investigated in Ref.~\cite{Yang:2020zrh}, where the mixing effect was not considered.

\item $P$-wave single charmed baryons of the $SU(3)$ flavor $\mathbf{6}_F$ was systematically investigated in Ref.~\cite{Yang:2021lce}, with the mixing effect taken into account.

\item $P$-wave single bottom baryons of the $SU(3)$ flavor $\mathbf{\bar3}_F$ was systematically investigated in Ref.~\cite{Wang:2024rai}, where the mixing effect was not considered.

\item $P$-wave single charmed baryons of the $SU(3)$ flavor $\mathbf{\bar3}_F$ will be systematically investigated in this study, with the mixing effect taken into account.

\end{itemize}
Note that the radiative decay properties of the $P$-wave charmed baryons belonging to the $SU(3)$ flavor $\mathbf{6}_F$ representation are still missing, which will be investigated in the near future.

This paper is organized as follows. In Sec.~\ref{secpcharmed} we briefly introduce our notation for the $P$-wave charmed baryons of the $SU(3)$ flavor $\mathbf{\bar3}_F$. In Sec.~\ref{sec:decay} and Sec.~\ref{sec:3FD} we calculate their strong and radiative decay properties, respectively. In Sec.~\ref{sec:mixing} we investigate the mixing effect between different HQET multiplets. In Sec.~\ref{secsummry} we discuss the results and conclude the paper.

\section{Notations and Input parameters}
\label{secpcharmed}

In this section we present a concise overview of the notations employed in our study. A charmed baryon is conceptualized as a bound state of one charmed quark ($c$) and two light quarks ($q = u, d, s$). When applying the Pauli principle to the two light quarks, it is required that their combined color, flavor, spin, and orbital degrees of freedom exhibit total antisymmetry:
\begin{itemize}

\item The color configuration of the two light quarks is antisymmetric, denoted as $\mathbf{\bar{3}}_C$ in the context of color $SU(3)$ symmetry.

\item The flavor configuration of the two light quarks can be either antisymmetric or symmetric, respectively denoted as $\mathbf{\bar{3}}_F$ and $\mathbf{6}_F$ in the context of flavor $SU(3)$ symmetry.

\item The spin configuration of the two light quarks can be either antisymmetric ($s_l = 0$) or symmetric ($s_l = 1$), where $s_l \equiv s_{qq}$ represents the spin of the two light quarks.

\item The orbital configuration of the two light quarks can be either antisymmetric ($\rho$-mode with $l_\rho = 1$ and $l_\lambda = 0$) or symmetric ($\lambda$-mode with $l_\rho = 0$ and $l_\lambda = 1$), where $l_\rho$ represents the orbital angular momentum between the two light quarks, and $l_\lambda$ represents the orbital angular momentum between the charm quark and the light quark system.

\end{itemize}
As illustrated in Fig.~\ref{fig:pwave}, the $P$-wave charmed baryons can be classified into eight distinct multiplets, four of which correspond to the $SU(3)$ flavor $\mathbf{\bar{3}}_F$ representation. These multiplets are characterized by the notation $[flavor, j_l, s_l, \rho/\lambda]$, where $j_l = l_\lambda \oplus l_\rho \oplus s_l$ represents the angular momentum of the light quark system. Within each multiplet, one or two baryons are identified, with their total angular momenta given by $J = j_l \oplus s_c = |j_l \pm 1/2|$.
\begin{table*}[ht]
\renewcommand{\arraystretch}{1.6}
\caption{Parameters of the $P$-wave charmed baryons belonging to the $SU(3)$ flavor $\mathbf{\bar{3}}_F$ representation.}
\label{tabmass}
\begin{tabular}{ c |c | c | c | c | c c | c | c}
\hline\hline
\multirow{2}{*}{doublets}&\multirow{2}{*}{~~B~~} & $\omega_c$ & ~Working region~ & ~~~~~~~$\overline{\Lambda}$~~~~~~~ & ~~~Baryon~~~ & ~~~~Mass~~~~~ & Difference & Decay constant
\\    &                                           & (GeV) & (GeV)      & (GeV)                &                               ($j^P$)       & (GeV)      & (MeV)        & (GeV$^{4}$)
\\ \hline\hline
\multirow{2}{*}{$[\mathbf{\bar 3}_F, 0, 1, \rho]$} &\multirow{1}{*}{$\Lambda_c$} & \multirow{1}{*}{$1.75\pm0.10$} & \multirow{1}{*}{$T=0.37/\rm CVG=65\%$} & \multirow{1}{*}{$0.99^{+0.09}_{-0.08}$} & $\Lambda_c(1/2^-)$ &$ 2.61^{+0.13}_{-0.13}$ & \multirow{1}{*}{-} & $0.020^{+0.006}_{-0.001}$ $(\Lambda^+_c(1/2^-))$
\\ \cline{2-9}
& \multirow{1}{*}{$\Xi_c$} & \multirow{1}{*}{$1.55\pm0.10$} & \multirow{1}{*}{$T=0.38/\rm CVG=75\%$} & \multirow{1}{*}{$1.14^{+0.04}_{-0.03}$} & $\Xi_c(1/2^-)$ & $2.81^{+0.06}_{-0.06}$& \multirow{1}{*}{-} &$0.027^{+0.001}_{-0.001}$ $(\Xi_c^0(1/2^-))$
\\ \hline
\multirow{4}{*}{$[\mathbf{\bar 3}_F, 1, 1, \rho]$} &\multirow{2}{*}{$\Lambda_c$} & \multirow{2}{*}{$1.54\pm0.10$} & \multirow{2}{*}{$0.27< T < 0.29$} & \multirow{2}{*}{$1.16^{+0.16}_{-0.13}$} & $\Lambda_c(1/2^-)$ & $2.59^{+0.19}_{-0.15}$ & \multirow{2}{*}{$50 \pm 9$} & $0.052 ^{+0.013}_{-0.009}~(\Lambda^+_c(1/2^-))$
\\ \cline{6-7}\cline{9-9}
&  & & & & $\Lambda_c(3/2^-)$ & $ 2.64^{+0.20}_{-0.14}$ & &$0.025 ^{+0.006}_{-0.004}~(\Lambda^+_c(3/2^-))$
\\ \cline{2-9}
& \multirow{2}{*}{$\Xi_c$} & \multirow{2}{*}{$1.79\pm0.10$} & \multirow{2}{*}{$0.27< T < 0.32$} & \multirow{2}{*}{$1.34 ^{+0.07}_{-0.05}$} & $\Xi_c(1/2^-)$ & $2.80^{+0.08}_{-0.08}$ & \multirow{2}{*}{$40 \pm 10$} & $0.079^{+0.008}_{-0.006} ~(\Xi_c^0(1/2^-))$
\\ \cline{6-7}\cline{9-9}
& & & & & $\Xi_c(3/2^-)$ & $2.84^{+0.08}_{-0.08}$ & &$0.037 ^{+0.004}_{-0.003}~(\Xi_c^0(3/2^-))$
\\ \hline

\multirow{4}{*}{$[\mathbf{\bar 3}_F, 2, 1, \rho]$} &\multirow{2}{*}{$\Lambda_c$} & \multirow{2}{*}{$1.85\pm0.10$} & \multirow{2}{*}{$T=0.31/\rm CVG=39\%$} & \multirow{2}{*}{$1.43^{+0.25}_{-0.20}$} & $\Lambda_c(3/2^-)$ &$2.65^{+0.24}_{-0.18}$ & \multirow{2}{*}{$90\pm9$}& \multirow{1}{*}{$0.072^{+0.027}_{-0.017}$$(\Lambda^+_c(3/2^-))$}
\\ \cline{6-7}\cline{9-9}
&  & & & & $\Lambda_c(5/2^-)$ &$2.74^{+0.20}_{-0.16} $&    &$0.031^{+0.012}_{-0.007}$$(\Lambda^+_c(5/2^-))$
\\ \cline{2-9}
 &\multirow{2}{*}{$\Xi_c$} & \multirow{2}{*}{$1.99\pm0.10$} & \multirow{2}{*}{$T=0.33/\rm CVG=35\%$} & \multirow{2}{*}{$1.50^{+0.12}_{-0.06}$} & $\Xi_c(3/2^-)$ &$2.85^{+0.15}_{-0.11}$ &\multirow{2}{*}{$80\pm5$} & \multirow{1}{*}{$0.089^{+0.008}_{-0.007}$
 $(\Xi^0_c(3/2^-))$}
\\ \cline{6-7}\cline{9-9}
&  & & & & $\Xi_c(5/2^-)$ &$2.93^{+0.10}_{-0.09}$ &    &$0.038^{+0.003}_{-0.003}$$(\Xi^0_c(5/2^-))$
\\ \hline

\multirow{4}{*}{$[\mathbf{\bar 3}_F,1,0,\lambda]$}& \multirow{2}{*}{ $\Lambda_c$} &\multirow{2}{*}{ $1.45\pm0.10$ }& \multirow{2}{*}{$T=0.30/\rm CVG=51\%$} &\multirow{2}{*}{ $0.96^{+0.15}_{-0.13} $} & $\Lambda_c(1/2^-)$ & $2.66 ^{+0.22}_{-0.19}$ &\multirow{2}{*}{ $30\pm5$}& $0.020 ^{+0.004}_{-0.003} ~(\Lambda^+_c(1/2^-))$
\\ \cline{6-7}\cline{9-9}
& & & & &$\Lambda_c(3/2^-)$ & $2.69^{+0.23}_{-0.19}$ & &$0.009 ^{+0.002}_{-0.001} ~(\Lambda^+_c(3/2^-))$
\\ \cline{2-9}
&\multirow{2}{*}{$\Xi_c$} & \multirow{2}{*}{$1.55\pm0.10$} & \multirow{2}{*}{$T=0.32/\rm CVG=50\%$} & \multirow{2}{*}{$1.06 ^{+0.07}_{-0.07}$} & $\Xi_c(1/2^-)$ & $2.79^{+0.09}_{-0.10}$ & \multirow{2}{*}{$30\pm 3$} & $0.025 ^{+0.002}_{-0.001}~(\Xi^0_c(1/2^-))$
\\ \cline{6-7}\cline{9-9}
& & & & & $\Xi_c(3/2^-)$ & $2.82 ^{+0.09}_{-0.10}$ & &$0.012 ^{+0.001}_{-0.001} ~(\Xi_c^0(3/2^-))$
\\ \hline\hline
\end{tabular}
\end{table*}

In Ref.~\cite{Chen:2017sci} we have conducted a comprehensive analysis of their mass spectra using the QCD sum rule method. We update these results and use them as input parameters for this study, as summarized in Table~\ref{tabmass}. Additionally, we need the following parameters for the ground-state charmed baryons and light pseudoscalar/vector mesons~\cite{PDG2024}:
\begin{itemize}

\item For the ground-state charmed baryons, we use
\begin{eqnarray}
\nonumber\Lambda_c(1/2^+)&:&\rm M=2286.46~{\rm MeV}\, ,
\\ \nonumber \Xi_c(1/2^+)&:&\rm M=2467.95~{\rm MeV}\, ,
\\ \nonumber \Sigma_c(1/2^+)&:&\rm M=2452.65~{\rm MeV}\, ,
\\ \nonumber \Sigma^*_c(3/2^+)&:&\rm M=2517.4~{\rm MeV}\, ,
\\ \nonumber \Xi^\prime_c(1/2^+)&:&\rm M=2578.2~{\rm MeV}\, ,
\\ \nonumber \Xi^*_c(3/2^+)&:&\rm M=2645.10~{\rm MeV}\, ,
\\ \nonumber \Omega_c(1/2^+)&:&\rm M=2695.2~{\rm MeV}\, ,
\\ \nonumber \Omega^*_c(1/2^+)&:&\rm M=2765.9~{\rm MeV}\, .
\end{eqnarray}

\item For the light pseudoscalar and vector mesons, we use
\begin{eqnarray}
\nonumber \pi(0^-)&:& m=138.04~{\rm MeV}\, ,
\\ \nonumber K(0^-) &:& m=495.65 ~{\rm MeV}\, ,
\\ \nonumber \rho(1^-)&:& m=775.21~{\rm MeV}\, ,
\\ \nonumber     &&\Gamma=148.2~{\rm MeV}\, ,
\\ \nonumber   &&g_{\rho\pi\pi}=5.94~{\rm GeV}^{-2}\, ,
\\ \nonumber K^{*}(1^-)&:& m=893.57 ~{\rm MeV}\, ,
\\ \nonumber &&\Gamma=49.1~{\rm MeV}\, ,
\\ \nonumber &&g_{K^* K\pi}=3.20~{\rm GeV}^{-2}\, .
\end{eqnarray}

\end{itemize}

\section{Strong decay properties}
\label{sec:decay}

In this section we investigate the strong decay properties of the $P$-wave charmed baryons belonging to the $SU(3)$ flavor $\mathbf{\bar3}_F$ representation. We focus on analyzing their decay patterns into ground-state charmed baryons accompanied by light pseudoscalar or vector mesons, including the following $S$-wave and $D$-wave decay channels:
\begin{widetext}
\begin{eqnarray}
&(a1)& {\bf \Gamma\Big[} \Lambda_c[1/2^-,1P] \rightarrow \Lambda_c + \pi {\Big]}
=  {\bf \Gamma\Big[}\Lambda_c^+[1/2^-,1P] \rightarrow \Lambda_c^+ +\pi^0 {\Big]} \, ,
\\ &(a2)& {\bf \Gamma\Big[} \Lambda_c[1/2^-,1P] \rightarrow \Sigma_c + \pi {\Big]}
=3\times {\bf \Gamma\Big[} \Lambda_c^+[1/2^-,1P] \rightarrow \Sigma_c^{++}+\pi^-\to\Lambda_c^++\pi^++\pi^- {\Big]} \, ,
\\ &(a3)& {\bf \Gamma\Big[} \Lambda_c[1/2^-,1P] \rightarrow \Sigma_c^* + \pi\rightarrow\Lambda_c+\pi+\pi {\Big]}
= 3 \times {\bf \Gamma \Big[}\Lambda_c^+[1/2^-,1P] \rightarrow \Sigma_c^{*++}+\pi^-\rightarrow \Lambda_c^++\pi^++\pi^-{\Big ]} \, ,
\\ &(a4)& {\bf\Gamma\Big[} \Lambda_c[1/2^-,1P] \rightarrow \Lambda_c + \rho \rightarrow\Lambda_c+\pi+\pi{\Big ]}
= {\bf \Gamma\Big[} \Lambda_c^+[1/2^-,1P] \rightarrow \Lambda_c^+ +\pi^++ \pi^- {\Big ]} \, ,
\\ &(a5)& { \bf\Gamma\Big[}\Lambda_c[1/2^-,1P] \rightarrow \Sigma_c + \rho\rightarrow\Sigma_c+\pi+\pi{\Big ]}
= 3 \times { \bf\Gamma\Big[}\Lambda_c^+[1/2^-,1P] \rightarrow \Sigma_c^+ +\pi^++ \pi^-{\Big ]} \, ,
\\ &(a6)&{\bf \Gamma\Big[}\Lambda_c[1/2^-,1P] \rightarrow \Sigma_c^* + \rho\rightarrow\Sigma_c^*+\pi+\pi{\Big ]}
= 3 \times { \bf\Gamma\Big[}\Lambda_c^+[1/2^-,1P] \rightarrow \Sigma_c^{*+} +\pi^++ \pi^-{\Big ]} \, ,
\\ &(b1)& {\bf \Gamma\Big[}\Lambda_c[3/2^-,1P] \rightarrow \Lambda_c + \pi{\Big ]}
= {\bf \Gamma\Big[}\Lambda_c^+[3/2^-] \rightarrow \Lambda_c^+ +\pi^0{\Big ]} \, ,
\\ &(b2)&{\bf \Gamma\Big[}\Lambda_c[3/2^-,1P] \rightarrow \Sigma_c + \pi\rightarrow\Lambda_c+\pi+\pi{\Big ]}
= 3 \times {\bf \Gamma\Big[}\Lambda_c^+[3/2^-,1P] \rightarrow \Sigma_c^{++} +\pi^-\rightarrow\Lambda_c^++\pi^++\pi^-{\Big ]} \, ,
\\ &(b3)&{\bf \Gamma\Big[}\Lambda_c[3/2^-,1P] \rightarrow \Sigma_c^* + \pi\rightarrow\Lambda_c+\pi+\pi{\Big ]}
= 3 \times {\bf \Gamma\Big[}\Lambda_c^+[3/2^-,1P] \rightarrow \Sigma_c^{*++} +\pi^-\rightarrow\Lambda_c^++\pi^+\pi^-{\Big ]} \, ,
\\ &(b4)&{\bf \Gamma\Big[} \Lambda_c[3/2^-,1P] \rightarrow \Lambda_c + \rho \rightarrow\Lambda_c+\pi+\pi{\Big ]}
= { \bf\Gamma\Big[}\Lambda_c^+[3/2^-,1P] \rightarrow \Lambda_c^+ +\pi^++ \pi^- {\Big ]} \, ,
\\ &(b5)& { \bf\Gamma\Big[}\Lambda_c[3/2^-,1P] \rightarrow \Sigma_c + \rho\rightarrow\Sigma_c+\pi+\pi{\Big ]}
= 3 \times { \bf\Gamma\Big[}\Lambda_c^+[3/2^-,1P] \rightarrow \Sigma_c^+ +\pi^++ \pi^-{\Big ]} \, ,
\\&(b6)& { \bf\Gamma\Big[}\Lambda_c[3/2^-,1P] \rightarrow \Sigma_c^* + \rho\rightarrow\Sigma_c^*+\pi+\pi {\Big ]}
= 3 \times {\bf \Gamma\Big[}\Lambda_c^0[3/2^-,1P] \rightarrow \Sigma_c^{*+} + \pi^++\pi^- {\Big ]} \, ,
\\ &(c1)& {\bf \Gamma\Big[}\Xi_c[1/2^-,1P] \rightarrow \Lambda_c + \bar K{\Big ]}
= {\bf \Gamma\Big[}\Xi_c^0[1/2^-,1P] \rightarrow\Lambda_c^+ +K^-{\Big ]} \, ,
\\ &(c2)&{\bf \Gamma\Big[}\Xi_c[1/2^-,1P] \rightarrow\Xi_c + \pi{\Big ]}
= {3\over2} \times {\bf \Gamma\Big[}\Xi_c^0[1/2^-,1P] \rightarrow \Xi_c^+ +\pi^-{\Big ]} \, ,
\\ &(c3)&{\bf \Gamma\Big[}\Xi_c[1/2^-,1P] \rightarrow\Sigma_c + \bar K{\Big ]}
= 3 \times {\bf \Gamma\Big[}\Xi_c^0[1/2^-,1P] \rightarrow \Sigma_c^+ +K^-{\Big ]} \, ,
\\ &(c4)& { \bf\Gamma\Big[}\Xi_c[1/2^-,1P] \rightarrow \Xi_c^{\prime}+\pi {\Big]}
= {3\over2}\times{\bf \Gamma\Big[}\Xi_c^0[1/2^-,1P]\rightarrow\Xi_c^{\prime+}+\pi^-{\Big]} \, ,
\\ &(c5)& {\bf \Gamma\Big[}\Xi_c[1/2^-,1P] \rightarrow \Sigma_c^* + K{\Big ]}
= 3 \times {\bf \Gamma\Big[}\Xi_c^0[1/2^-,1P] \rightarrow \Sigma_c^{*+} + K^-{\Big ]} \, ,
\\ &(c6)& {\bf \Gamma\Big[}\Xi_c[1/2^-,1P] \rightarrow \Xi_c^* + \pi\rightarrow \Xi_c+\pi+\pi {\Big ]}
= {9 \over 2} \times {\bf \Gamma\Big[}\Xi_c^{0}[1/2^-,1P] \rightarrow\Xi_c^{*+} + \pi^-\rightarrow\Xi_c^++\pi^0+\pi^-{\Big]} \, ,
\\ &(c7)& {\bf \Gamma\Big[}\Xi_c[1/2^-,1P] \rightarrow \Lambda_c + \bar{K}^*\rightarrow\Lambda_c+\bar K+\pi{\Big ]}
=3\times  {\bf \Gamma\Big[}\Xi_c^0[1/2^-,1P] \rightarrow \Lambda_c^+ +  K^- +\pi^0{\Big ]} \, ,
\\ &(c8)& {\bf \Gamma\Big[}\Xi_c[1/2^-,1P] \rightarrow\Xi_c + \rho\rightarrow\Xi_c+\pi+\pi{\Big ]}
= {3\over2} \times {\bf \Gamma\Big[}\Xi_c^0[1/2^-,1P] \rightarrow \Xi_c^+ + \pi^0+\pi^-{\Big ]} \, ,
\\ &(c9)& {\bf \Gamma\Big[}\Xi_c[1/2^-,1P] \rightarrow \Sigma_c^* + \bar K^*\rightarrow\Sigma_c^{*}+\bar K+\pi{\Big ]}
= 9 \times {\bf \Gamma\Big[}\Xi_c^0[1/2^-,1P] \rightarrow \Sigma_c^{*+} + K^-+ \pi^0{\Big ]} \, ,
\\ &(c10)& {\bf \Gamma\Big[}\Xi_c[1/2^-,1P] \rightarrow \Xi_c^{*}+ \rho\rightarrow\Xi_c^{*}+\pi+\pi {\Big ]}
={3\over2} \times {\bf \Gamma\Big[}\Xi_c^0[1/2^-,1P] \rightarrow \Xi_c^{*+} + \pi^0+\pi^- {\Big ]}\, ,
\\ &(d1)& {\bf \Gamma\Big[}\Xi_c[3/2^-,1P] \rightarrow \Lambda_b + \bar K{\Big ]}
= {\bf \Gamma\Big[}\Xi_c^0[3/2^-,1P] \rightarrow\Lambda_c^+ +K^-{\Big ]} \, ,
\\ &(d2)&{\bf \Gamma\Big[}\Xi_c[3/2^-,1P] \rightarrow\Xi_c + \pi{\Big ]}
= {3\over2} \times {\bf \Gamma\Big[}\Xi_c^0[3/2^-,1P] \rightarrow \Xi_c^+ +\pi^-{\Big ]} \, ,
\\ &(d3)&{\bf \Gamma\Big[}\Xi_c[3/2^-,1P] \rightarrow \Sigma_c + \bar K{\Big ]}
= 3 \times {\bf \Gamma\Big[}\Xi_c^0[3/2^-,1P] \rightarrow \Sigma_c^+ +K^-{\Big ]} \, ,
\\ &(d4)& { \bf\Gamma\Big[}\Xi_c[3/2^-,1P] \rightarrow \Xi_c^{\prime}+\pi {\Big]}
= {3\over2}\times{\bf \Gamma\Big[}\Xi_c^0[3/2^-,1P]\rightarrow\Xi_c^{\prime+}+\pi^-{\Big]} \, ,
\\ &(d5)& {\bf \Gamma\Big[}\Xi_c[3/2^-,1P] \rightarrow \Sigma_c^* + \bar K{\Big ]}
= 3 \times {\bf \Gamma\Big[}\Xi_c^0[3/2^-,1P] \rightarrow \Sigma_c^{*+} + K^-{\Big ]} \, ,
\\ &(d6)& {\bf \Gamma\Big[}\Xi_c[3/2^-,1P] \rightarrow \Xi_c^* + \pi\rightarrow \Xi_c+\pi+\pi {\Big ]}
= {9 \over 2} \times {\bf \Gamma\Big[}\Xi_c^{*0}[3/2^-,1P] \rightarrow\Xi_c^{*+} + \pi^-\rightarrow\Xi_c^++\pi^0+\pi^-{\Big]} \, ,
\\ &(d7)& {\bf \Gamma\Big[}\Xi_c[3/2^-,1P] \rightarrow \Lambda_c + \bar{K}^*\rightarrow\Lambda_b+\bar K+\pi{\Big ]}
=3\times  {\bf \Gamma\Big[}\Xi_c^0[3/2^-,1P] \rightarrow \Lambda_c^+ +  K^- +\pi^0{\Big ]} \, ,
\\ &(d8)& {\bf \Gamma\Big[}\Xi_c[3/2^-,1P] \rightarrow \Xi_c + \rho\rightarrow\Xi_b+\pi+\pi{\Big ]}
= {3\over2} \times {\bf \Gamma\Big[}\Xi_c^0[3/2^-,1P] \rightarrow \Xi_c^+ + \pi^0+\pi^-{\Big ]} \, ,
\\ &(d9)& {\bf \Gamma\Big[}\Xi_c[3/2^-,1P] \rightarrow \Sigma_c^* + \bar K^*\rightarrow\Sigma_c^{*0}+\bar K+\pi{\Big ]}
= 9 \times {\bf \Gamma\Big[}\Xi_c^0[3/2^-,1P] \rightarrow \Sigma_c^{*+} + K^-+ \pi^0{\Big ]} \, ,
\\ &(d10)& {\bf \Gamma\Big[} \Xi_c[3/2^-,1P] \rightarrow \Xi_c^{*}+ \rho\rightarrow\Xi_c^{*}+\pi+\pi {\Big ]}
={3\over2} \times {\bf \Gamma\Big[}\Xi_c^0[3/2^-,1P] \rightarrow \Xi_c^{*+} + \pi^0+\pi^- {\Big ]}\, .
\label{eq:couple}
\end{eqnarray}
\end{widetext}
Their partial decay widths can be evaluated through the following Lagrangians:
\begin{eqnarray}
&&\mathcal{L}^S_{X_c({1/2}^-) \rightarrow Y_c({1/2}^+) P}
\\ \nonumber&& ~~~~~~~~~~~~\, = g^S {\bar X_c}(1/2^-) Y_c(1/2^+) P \, ,
\\ &&\mathcal{L}^S_{X_c({3/2}^-) \rightarrow Y_c({3/2}^+) P}
\\ \nonumber&& ~~~~~~~~~~~~\, = g^S {\bar X_{c\mu}}(3/2^-)Y_c^{\mu}(3/2^+) P \, ,
\\ &&\mathcal{L}^S_{X_c({1/2}^-) \rightarrow Y_c({1/2}^+) V}
\\ \nonumber&& ~~~~~~~~~~~~\, = g^S {\bar X_c}(1/2^-) \gamma_\mu \gamma_5 Y_c(1/2^+) V^\mu \, ,
\\ &&\mathcal{L}^S_{X_c({1/2}^-) \rightarrow Y_c({3/2}^+) V}
\\ \nonumber&& ~~~~~~~~~~~~\, = g^S {\bar X_{c}}(1/2^-) Y_{c}^{\mu}(3/2^+) V_\mu \, ,
\\ &&\mathcal{L}^S_{X_c({3/2}^-) \rightarrow Y_c({1/2}^+) V}
\\ \nonumber&& ~~~~~~~~~~~~\, = g^S {\bar X_{c}^{\mu}}(3/2^-) Y_{c}(1/2^+) V_\mu \, ,
\\ &&\mathcal{L}^S_{X_c({3/2}^-) \rightarrow Y_c({3/2}^+) V}
\\ \nonumber&& ~~~~~~~~~~~~\, = g^S {\bar X_c}^{\nu}(3/2^-) \gamma_\mu \gamma_5 Y_{c\nu}(3/2^+) V^\mu \, ,
\\&& \mathcal{L}^S_{X_c({5/2}^-) \rightarrow Y_c({3/2}^+) V}
\\ \nonumber&& ~~~~~~~~~~~~\, = g^S {\bar X_{c}^{\mu\nu}}(5/2^-) Y_{c\mu}(3/2^+) V_\nu
\\ \nonumber &&~~~~~~~~~~~~\, + g^S {\bar X_{c}^{\nu\mu}}(5/2^-) Y_{c\mu}(3/2^+) V_\nu \, ,
\\ && \mathcal{L}^D_{X_c({1/2}^-) \rightarrow Y_c({3/2}^+) P}
\\ \nonumber && ~~~~~~~~~~~\, = g^D {\bar X_c}(1/2^-) \gamma_\mu \gamma_5 Y_{c\nu}(3/2^+) \partial^{\mu} \partial^{\nu}P \, ,
\\ && \mathcal{L}^D_{X_c({3/2}^-) \rightarrow Y_c({1/2}^+) P}
\\ \nonumber && ~~~~~~~~~~~\, = g^D {\bar X_{c\mu}}(3/2^-) \gamma_\nu \gamma_5 Y_{c}(1/2^+) \partial^{\mu} \partial^{\nu}P \, ,
\label{eq:lagrangians}
\\ && \mathcal{L}^D_{X_c({3/2}^-) \rightarrow Y_c({3/2}^+) P}
\\ \nonumber && ~~~~~~~~~~~\, = g^D {\bar X_{c\mu}}(3/2^-) Y_{c\nu}(3/2^+) \partial^{\mu} \partial^{\nu}P \, ,
\\ && \mathcal{L}^D_{X_c({5/2}^-) \rightarrow Y_c({1/2}^+) P}
\\ \nonumber && ~~~~~~~~~~~\, = g^D {\bar X_{c\mu\nu}}(5/2^-) Y_{c}(1/2^+) \partial^{\mu} \partial^{\nu}P \, ,
\\ && \mathcal{L}^D_{X_c({5/2}^-) \rightarrow Y_c({3/2}^+) P}
\\ \nonumber && ~~~~~~~~~~~\, = g^D {\bar X_{c\mu\nu}}(5/2^-) \gamma_\rho \gamma_5 Y_{c}^{\mu}(3/2^+) \partial^{\nu} \partial^{\rho}P
\\ \nonumber && ~~~~~~~~~~~\, + g^D {\bar X_{c\mu\nu}}(5/2^-) \gamma_\rho \gamma_5 Y_{c}^{\nu}(3/2^+) \partial^{\mu} \partial^{\rho}P \, ,
\end{eqnarray}
where $S$ and $D$ represent the $S$-wave and $D$-wave decay channels, respectively. The fields $X_c^{(\mu\nu)}$ and $Y_c^{(\mu)}$ represent the $P$-wave charmed baryons and the ground-state charmed baryons, respectively.

We use the $P$-wave charmed baryon $\Lambda_c^+({3/2}^-)$ belonging to the  $[\mathbf{\bar 3}_F, 1, 1, \rho]$ doublet as a representative case, and calculate its $D$-wave decay into the final state $\Sigma_c^{++}\pi^-$. To accomplish this, we consider the two-point correlation function:
\begin{eqnarray}
&& \Pi^{\alpha}(\omega, \omega^\prime)
\\ \nonumber &=& \int d^4 x~e^{-i k \cdot x}~\langle 0 | J^\alpha_{3/2,-,\Lambda_c^+,1,1,\rho}(0) \bar J_{\Sigma_c^{++}}(x) | \pi^-(q) \rangle
\\ \nonumber &=& {1+v\!\!\!\slash\over2} G^{\alpha}_{\Lambda_c^+[{3\over2}^-] \rightarrow   \Sigma_c^{++}\pi^-} (\omega, \omega^\prime) \, ,
\end{eqnarray}
where $k^\prime = k + q$, $\omega = v \cdot k$, and $\omega^\prime = v \cdot k^\prime$, with $k^\prime_\mu$, $k_\mu$, and $q_\mu$ the four-momenta of the initial baryon, the final baryon, and the light meson, respectively.

\begin{widetext}
At the hadron level, we write
$G^{\alpha}_{\Lambda_c^+[{3\over2}^-] \rightarrow \Sigma_c^{++}\pi^-}$ as ($\cdots$ contains other possible amplitudes):
\begin{eqnarray}
G^{\alpha}_{\Lambda_c^+[{3\over2}^-] \rightarrow \Sigma_c^{++}\pi^-} (\omega, \omega^\prime)
\label{G0C1}
= g^D_{\Lambda_c^+[{3\over2}^-] \rightarrow \Sigma_c^{++}\pi^-} {  f_{\Lambda_c^+[{3\over2}^-]} f_{\Sigma_c^{+}} \over (\bar \Lambda_{\Lambda_c^+[{3\over2}^-]} - \omega^\prime) (\bar \Lambda_{\Sigma_c^{++}} - \omega)} \gamma_{\alpha}\gamma_5+ \cdots \, .
\end{eqnarray}
At the quark-gluon level, we calculate $G^\alpha_{\Lambda_c^+[{3\over2}^-] \rightarrow \Sigma_c^{++}\pi^-}$ using the method of operator product expansion (OPE):
\begin{eqnarray}
\label{eq:g1}
&& G^{\alpha}_{\Lambda_c^+[{3\over2}^-] \rightarrow \Sigma_c^{++}\pi^-} (\omega, \omega^\prime)
\\ \nonumber &=& \int_0^\infty dt \int_0^1 du e^{i (1-u) \omega^\prime t} e^{i u \omega t}\times4 \times \Bigg(-\frac{{{\rm{}}{f_\pi u_0}}}{{4{\pi ^2t^3}}}{\phi _{2;\pi }}+\frac{ f_{\pi}{u_0}}{64 \pi^2t}\phi_{4;\pi}(u_0)+\frac{ f_{\pi}m^2_{\pi}u_0t}{144(m_{u})}\langle\bar q q\rangle\phi _3^\sigma(u_0)
\\ \nonumber
&-&\frac{{{f_\pi }m_\pi ^2{u_0t^3}}}{{2304({m_u})}}\langle {g_s}\sigma Gs\rangle \phi _3^\sigma ({u_0}) \Bigg)\times\gamma_{\alpha}\gamma_5 +\int_0^\infty dt \int_0^1 du \int \mathcal{D} \underline{\alpha} e^{i \omega^{\prime} t(\alpha_2 + u \alpha_3)} e^{i \omega t(1 - \alpha_2 - u \alpha_3)}\times \frac{1}{2} \Bigg( (\frac{{{f_\pi }{\Phi _{4;\pi }}(\underline\alpha){\alpha _3}{u_0}}}{{24{\pi ^2t}}}
\\ \nonumber
&+& \frac{{{f_\pi }{\Phi _{4;\pi }}(\underline\alpha){\alpha _2}{u_0}}}{{24{\pi ^2t}}}
+ \frac{{{f_\pi }{\Phi _{4;\pi }}(\underline\alpha){\alpha _3}{u_0}}}{{48{\pi ^2t}}}+\frac{{{f_\pi }{\widetilde\Phi _{4;\pi }}(\underline\alpha){\alpha _3}{u_0}}}{{48{\pi ^2t}}}- \frac{{{f_\pi }{\Phi _{4;\pi }}(\underline\alpha){u_0}}}{{24{\pi ^2t}}} + \frac{{{f_\pi }{\Phi _{4;\pi }}(\underline\alpha){\alpha _2}}}{{48{\pi ^2t}}}
+ \frac{{{f_\pi }{\widetilde\Phi _{4;\pi }}(\underline\alpha){\alpha _2}}}{{48{\pi ^2t}}}
\\ \nonumber
&-&\frac{{{f_\pi }{\Phi _{4;\pi }}(\underline\alpha)}}{{48{\pi ^2t}}}
 -\frac{{{f_\pi }{\widetilde\Phi _{4;\pi }}(\underline\alpha)}}{{48{\pi ^2t}}})
-(\frac{{{f_\pi }{\Phi _{4;\pi }}(\underline\alpha)u_0}}{{12{\pi ^2t^2(v\cdot q)}}} + \frac{{{f_\pi }\widetilde\Phi _{4;\pi }(\underline\alpha){u_0}}}{{4{\pi ^2t^2(v\cdot q)}}} + \frac{{{f_\pi }\Psi_{4;\pi}(\underline\alpha){u_0}}}{{24{\pi ^2t^2(v\cdot q)}}}
  + \frac{{{f_\pi }\widetilde\Psi _{4;\pi }(\underline\alpha){u_0}}}{{8{\pi ^2t^2(v\cdot q)}}}
 + \frac{{{f_\pi }{\Phi_{4;\pi }}(\underline\alpha)}}{{8{\pi ^2t^2(v\cdot q)}}}
  \\  \nonumber
  &+& \frac{{{f_\pi }\widetilde\Phi _{4;\pi }}(\underline\alpha)}{{24{\pi ^2t^2(v\cdot q)}}}
+ \frac{{{f_\pi }{\Phi _{4;\pi }}(\underline\alpha)}}{{24{\pi ^2t^2(v\cdot q)}}}- \frac{{{f_\pi } \widetilde\Phi _{4;\pi }(\underline\alpha)}}{{24{\pi ^2t^2(v\cdot q)}}})\Bigg)\times\gamma_{\alpha}\gamma_5 \, .
\end{eqnarray}
 Then we perform the Borel transformation to both Eq.~(\ref{G0C1}) at the hadron level and Eq.~(\ref{eq:g1}) at the quark-gluon level:
\begin{eqnarray}
\label{eq:g}
&& g^D_{\Lambda_c^+[{3\over2}^-] \rightarrow\Sigma_c^{++}\pi^-} f_{\Lambda_c^+[{3\over2}^-]}
f_{\Sigma_c^{++}} e^{- {\bar \Lambda_{\Lambda_c^+[{3\over2}^-]} \over T_1}} e^{ - {\bar \Lambda_{\Sigma_c^{++}} \over T_2}}
\\ \nonumber &=& 4 \times \Bigg(-\frac{{{\rm{}}{f_\pi u_0 }}}{{4{\pi ^2}}}{T^4}{f_3}(\frac{{{\omega _c}}}{T}){\phi _{2;\pi }}({u_0})|_{u_0 = \frac{1}{2}}+\frac{ f_{\pi}{u_0}}{64 \pi^2}T^2f_1(\frac{\omega_c}{T})\phi_{4;\pi}(u_0)u_0|_{u_0 = \frac{1}{2}}+\frac{ f_{\pi}m^2_{\pi}u_0}{144(m_{u})}\langle\bar q q\rangle\phi _3^\sigma(u_0)|_{u_0 = \frac{1}{2}}
\\ \nonumber
&-&\frac{{{f_\pi }m_\pi ^2{u_0}}}{{2304({m_u})}}{T^{-2}}\langle {g_s}\sigma Gs\rangle \phi _3^\sigma ({u_0})|{_{{u_0} = \frac{1}{2}}} \Bigg) + \frac{1}{2} \times f_1(\frac{{{\omega _c}}}{T})\int_0^{\frac{1}{2}} d {\alpha _2}\int_{\frac{1}{2} - {\alpha_2}}^{1 - {\alpha _2}}  \frac{{{{T}^2}}}{{\alpha 3}}\Bigg(\frac{{{f_\pi }{\Phi _{4;\pi }}(\underline\alpha){\alpha _3}{u_0}}}{{24{\pi ^2}}} + \frac{{{f_\pi }{\Phi _{4;\pi }}(\underline\alpha){\alpha _2}{u_0}}}{{24{\pi ^2}}}
\\ \nonumber
 &+& \frac{{{f_\pi }{\Phi _{4;\pi }}(\underline\alpha){\alpha _3}{u_0}}}{{48{\pi ^2}}}+\frac{{{f_\pi }{\widetilde\Phi _{4;\pi }}(\underline\alpha){\alpha _3}{u_0}}}{{48{\pi ^2}}}- \frac{{{f_\pi }{\Phi _{4;\pi }}(\underline\alpha){u_0}}}{{24{\pi ^2}}} + \frac{{{f_\pi }{\Phi _{4;\pi }}(\underline\alpha){\alpha _2}}}{{48{\pi ^2}}}+ \frac{{{f_\pi }{\widetilde\Phi _{4;\pi }}(\underline\alpha){\alpha _2}}}{{48{\pi ^2}}}-\frac{{{f_\pi }{\Phi _{4;\pi }}(\underline\alpha)}}{{48{\pi ^2}}}
 -\frac{{{f_\pi }{\widetilde\Phi _{4;\pi }}(\underline\alpha)}}{{48{\pi ^2}}}\Bigg)
 \\ \nonumber
 &-&\frac{1}{2}\times{f_1}(\frac{{{\omega _c}}}{T})\int_0^{\frac{1}{2}} d {\alpha _2}\int_{\frac{1}{2} - {\alpha _2}}^{1 - {\alpha _2}} d {\alpha _3}\frac{{{u_0}{T^2}}}{{\alpha_4}}\Bigg(\frac{{{f_\pi }{\Phi _{4;\pi }}(\underline\alpha)u_0}}{{12{\pi ^2}}} + \frac{{{f_\pi }\widetilde\Phi _{4;\pi }(\underline\alpha){u_0}}}{{4{\pi ^2}}} + \frac{{{f_\pi }\Psi_{4;\pi}(\underline\alpha){u_0}}}{{24{\pi ^2}}}
  + \frac{{{f_\pi }\widetilde\Psi _{4;\pi }(\underline\alpha){u_0}}}{{8{\pi ^2}}}
 \\  \nonumber
 &+& \frac{{{f_\pi }{\Phi_{4;\pi }}(\underline\alpha)}}{{8{\pi ^2}}}+ \frac{{{f_\pi }\widetilde\Phi _{4;\pi }}(\underline\alpha)}{{24{\pi ^2}}}+ \frac{{{f_\pi }{\Phi _{4;\pi }}(\underline\alpha)}}{{24{\pi ^2}}}- \frac{{{f_\pi } \widetilde\Phi _{4};\pi }(\underline\alpha)}{{24{\pi ^2}}}\Bigg) \, .
\end{eqnarray}
\end{widetext}
In the above formalism, $\alpha = \{ \alpha_1, \alpha_2, \alpha_3 \}$ and $\int \mathcal{D}\underline{\alpha} = \int_0^1 d\alpha_1 \int_0^1 d\alpha_2 \int_0^1 d\alpha_3$; $\Phi_{\cdots}(\underline\alpha)$ in Eq.~(\ref{eq:g1}) contains the constraint $\delta(\alpha_1 + \alpha_2 + \alpha_3 - 1)$, while this $\delta$ function has been integrated out in Eq.~(\ref{eq:g}); the function $f_n(x)$ is defined as $f_n(x) \equiv 1 - e^{-x} \sum_{k=0}^n {x^k \over k!}$. The parameters  $\omega$ and $\omega^\prime$  are transformed into $T_1$ and $T_2$, respectively. We adopt the symmetric point $T_1 = T_2 = 2T$, which yields  $u_0 = {T_1 \over T_1 + T_2} = {1\over2}$. The threshold value $\omega_c$ is set to 1.54 GeV, corresponding to the averaged threshold value derived from the mass sum rules of $\Lambda_c^{+}({3/2}^-)$ and $\Sigma_c^{++}(1/2^+)$. The Borel mass $T$ is chosen within the range $0.27~\rm{GeV}<T<0.29~\rm{GeV}$, derived from the mass sum rules of $\Lambda_c^{+}({3/2}^-)$. The light-cone distribution amplitudes incorporated in the above sum-rule expressions are taken from Refs.~\cite{Ball:1998je,Ball:2006wn,Ball:2004rg,Ball:1998kk,Ball:1998sk,Ball:1998ff,Ball:2007rt,Ball:2007zt,Ball:2002ps}.

\begin{figure}[hbt]
\begin{center}
\scalebox{0.9}{\includegraphics{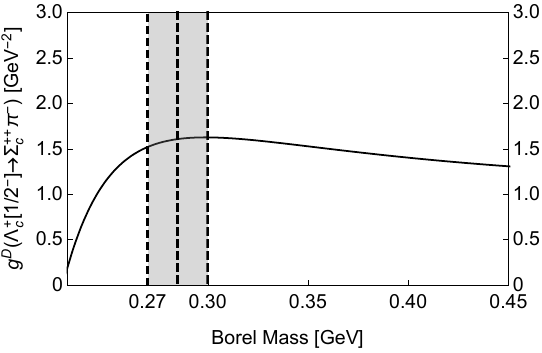}}
\caption{The coupling constant $g^D_{\Lambda^+_c[\frac{3}{2}^-]\to\Sigma_c^{++}\pi^-}$ as a function of the Borel mass $T$. Here the charmed baryon $\Lambda_c^+({3/2}^-)$ belonging to the  $[\mathbf{\bar 3}_F, 1, 1, \rho]$ doublet is investigated.}
\label{fig:311R}
\end{center}
\end{figure}

For the current analysis, we employ the following values for the condensates at the renormalization scale of 1~GeV~\cite{Yang:1993bp,Hwang:1994vp,Ovchinnikov:1988gk,Narison:2002woh,Jamin:2002ev,Ioffe:2002be,Shifman:2001ck,Gimenez:2005nt}:
%
\begin{eqnarray}
\nonumber && \langle \bar qq \rangle = - (0.24 \pm 0.01)^3 \mbox{ GeV}^3 \, ,
\\ \nonumber && \langle \bar ss \rangle = (0.8\pm 0.1)\times \langle\bar qq \rangle \, ,
\\ && \langle g_s \bar q \sigma G q \rangle = M_0^2 \times \langle \bar qq \rangle\, ,
\label{eq:condensates1}
\\ \nonumber && \langle g_s \bar s \sigma G s \rangle = M_0^2 \times \langle \bar ss \rangle\, ,
\\ \nonumber && M_0^2= 0.8 \mbox{ GeV}^2\, ,
\\ \nonumber && \langle g_s^2GG\rangle =(0.48\pm 0.14) \mbox{ GeV}^4\, .
\end{eqnarray}
As illustrated in Fig.~\ref{fig:311R}, the coupling constant is derived from Eq.~(\ref{eq:g}) as
\begin{eqnarray}
\nonumber g^D_{\Lambda_c^+[{3\over2}^-] \rightarrow \Sigma_c^{++}\pi^-}
 &=&1.58^{+0.76}_{-0.89}{~\rm GeV}^{-2} \, ,
\end{eqnarray}
where the associated uncertainties originate from four primary sources: (1) the variation in the Borel mass, (2) the input parameters of the $\Sigma_c^{++}(1/2^+)$ state, (3) the input parameters of the $\Lambda_c^+({3/2}^-)$ state, and (4) the QCD parameters specified in Eq.~(\ref{eq:condensates1}).

The relevant decay width can be calculated through
\begin{eqnarray}
&&\Gamma^D\left(\Lambda_c(3/2^-)\to\Sigma_c+\pi\right)\\
\nonumber &\equiv&3 \times\Gamma^D\left(\Lambda_c^+(3/2^-)\to\Sigma_c^{++}+\pi^-\right)\\
\nonumber &=& {3|\vec{p_2 }| g_{0\to 1+2}^2\over 32\pi m_0^2} \times \rm Tr[(p_0\!\!\!\slash+m_0)\gamma_5\gamma_{\beta1}(p_1\!\!\!\slash+m_1)\gamma_{\beta2}\gamma_5]
\\ \nonumber
&\times&({g_{{\alpha _1}{\alpha _2}}} - \frac{1}{3}{\gamma _{{\alpha _2}}}{\gamma _{{\alpha _1}}} - \frac{{{p_0^{{\alpha _2}}}{\gamma _{{\alpha _1}}} - {p_0^{{\alpha _1}}}{\gamma _{{\alpha _2}}}}}{{3m_0}} - \frac{{2{p_0^{{\alpha _2}}}{p_0^{{\alpha _1}}}}}{{3{m_0^2}}})
\\ \nonumber
&\times& p_2^{\alpha_1}p_2^{\alpha_2}p_2^{\beta_1}p_2^{\beta_2}\, ,
\end{eqnarray}
where $0$, $1$, and $2$ denote $\Lambda_c^+(3/2^-)$, $\Sigma_c^{++}(1/2^+)$, and $\pi^-$, respectively. Numerically, we obtain
\begin{eqnarray}
\Gamma^D_{\Lambda_c^+[{3\over2}^-] \rightarrow \Sigma_c\pi }&=& 4.0^{+4.8}_{-4.0}\times10^{-3}{\rm~MeV} \, .
\end{eqnarray}

In a similar manner, we investigate the four $SU(3)$ flavor $\mathbf{\bar 3}_F$ charmed baryon multiplets: $[\mathbf{\bar 3}_F, 1, 0, \lambda]$, $[\mathbf{\bar 3}_F, 0, 1, \rho]$, $[\mathbf{\bar 3}_F, 1, 1, \rho]$, and $[\mathbf{\bar 3}_F, 2, 1, \rho]$. The obtained results are summarized in Table~\ref{tab:decayb3f}.

\begin{table*}[hbt]
\begin{center}
\renewcommand{\arraystretch}{1.4}
\caption{Strong decay properties of the $P$-wave charmed baryons belonging to the $SU(3)$ flavor $\mathbf{\bar 3}_F$ representation.}
\setlength{\tabcolsep}{0.1mm}{
\begin{tabular}{c| c | c | c | c | c | c | c | c}
\hline\hline
\multirow{2}{*}{Multiplets}& ~B~ & ~~Mass~~ & Difference & \multirow{2}{*}{~~~~~Decay channels~~~~~}  & ~~$g^{S/D}$~~  & ~Decay width~ & ~Total~ & \multirow{2}{*}{Candidate}
\\ & ($j^P$) & ({GeV})& ({MeV}) & & ($\rm GeV^{-2}$) & ({MeV}) & ({MeV})
\\ \hline\hline
\multirow{12}{*}{$[\mathbf{\bar 3}_F,1,0,\lambda]$}&\multirow{3}{*}{$\Lambda_c({1\over2}^-)$}&\multirow{3}{*}{$2.66^{+0.22}_{-0.19}$}
&\multirow{6}{*}{$30\pm5$}&$\Lambda_c({1\over2}^-)\to \Sigma_c\pi\to\Lambda_c\pi\pi$ &$g^S=17.0^{+7.1}_{-7.2}$&$0\sim15000$&
&\multirow{3}{*}{--}
\\ \cline{5-7}
&&&&$\Lambda_c({1\over2}^-)\to \Sigma_c^*\pi\to\Lambda_c\pi\pi$&$g^D=4.3^{+0.8}_{-1.1}$&$1^{+7}_{-1}\times 10^{-5}$&\multirow{1}{*}{$0\sim15000$}&
\\ \cline{5-7}
&&&&$\Lambda_c({1\over2}^-)\to \Lambda_c\rho\to\Lambda_c\pi\pi$&$g^S=1.10^{+0.17}_{-0.28}$&$3.1^{+14.2}_{-~1.7}\times10^{-2}$&&
\\ \cline{2-3} \cline{5-9}
&\multirow{3}{*}{$\Lambda_c({3\over2}^-)$}&\multirow{3}{*}{$2.69^{+0.23}_{-0.19}$}&
&\multirow{2}{*}{$\Lambda_c({3\over2}^-)\to\Sigma_c^*\pi\to\Lambda_c\pi\pi$}&$g^S=8.2^{+2.7}_{-3.0}$
&$0\sim4800$ &\multirow{4}{*}{$0\sim4800$}
\\&&&&&$g^D=3.0^{+0.6}_{-0.6}$&$3^{+10}_{-~3}$& &\multirow{2}{*}{--}
\\ \cline{5-7}
&&&&$\Lambda_c({3\over2}^-)\to\Sigma_c\pi$&$g^D=5.2^{+1.2}_{-1.0}$&$10^{+40}_{-10}$&&
\\ \cline{5-7}
&&&&$\Lambda_c({3\over2}^-)\to \Lambda_c\rho\to\Lambda_c\pi \pi$ &$g^S=1.72^{+0.39}_{-0.36}$ &$6.4^{+24.2}_{-~4.0}\times10^{-2}$ &&
\\ \cline{2-9}
&\multirow{3}{*}{$\Xi_c({1\over2}^-)$}&\multirow{3}{*}{$2.79^{+0.09}_{-0.10}$}
&\multirow{6}{*}{$30\pm3$}& $\Xi_c({1\over2}^-)\to \Xi_c^{\prime}\pi$&$g^S=14.7^{+6.7}_{-9.3}$&$0\sim16000$& &\multirow{3}{*}{--}
\\ \cline{5-7}
&&&&$\Xi_c({1\over2}^-)\to\Xi_c^*\pi\to\Xi_c\pi\pi$&$g^D=3.2^{+1.0}_{-1.0}$&$3^{+8}_{-3}\times 10^{-8}$&\multirow{1}{*}{$0\sim16000$}&
\\ \cline{5-7}
&&&&$\Xi_c({1\over2}^-)\to\Xi_c\rho\to\Xi_c\pi\pi$&$g^S=1.26^{+0.34}_{-0.32}$
&$2.7^{+4.4}_{-1.4}\times10^{-2}$&
\\ \cline{2-3} \cline{5-9}
&\multirow{3}{*}{$\Xi_c({3\over2}^-)$}&\multirow{3}{*}{$2.82^{+0.09}_{-0.10}$}&
&$\Xi_c({3\over2}^-)\to\Xi_ c^{\prime}\pi$&$g^D=3.9^{+1.1}_{-1.1}$&$3^{+3}_{-1}$& &\multirow{5}{*}{--}
\\ \cline{5-7}
&&&&\multirow{2}{*}{$\Xi_c({3\over2}^-)\to\Xi_c^{*}\pi$}&$g^S=6.9^{+2.8}_
{-3.8}$&$1100^{+1100}_{-~950} $&\multirow{2}{*}{$1100^{+1100}_{-~950}$}&
\\&&&&&$g^D=2.3^{+0.7}_{-0.7}$&$2^{+7}_{-2}\times10^{-3}$&
\\ \cline{5-7}
&&&&$\Xi_c({3\over2}^-)\to\Xi_c \rho\to\Xi_c \pi \pi$&$g^S=1.77^{+0.48}_{-0.44}$&$1.0^{+1.7}_{-0.6}\times10^{-2}$&
\\ \hline\hline
\multirow{2}{*}{$[\mathbf{\bar 3}_F,0,1,\rho]$}&\multirow{1}{*}{$\Lambda_c({1\over2}^-)$}&\multirow{1}{*}{$2.61^{+0.13}_{-0.13}$}
&--&--&--&--&--&--
\\ \cline{2-9}
&\multirow{1}{*}{$\Xi_c({1\over2}^-)$}&\multirow{1}{*}{$2.81^{+0.06}_{-0.06}$}
&\multirow{1}{*}{--}&\multirow{1}{*}{$\Xi_c({1\over2}^-)\to \Xi_c\pi$}
&\multirow{1}{*}{$g^S=9.6^{+2.5}_{-3.8}$}&\multirow{1}{*}{$5700^{+3600}_{-3800}$}&
\multirow{1}{*}{$5700^{+3600}_{-3800}$}&--
\\ \hline\hline
\multirow{12}{*}{$[\mathbf{\bar 3}_F,1,1,\rho]$}&\multirow{3}{*}{$\Lambda_c({1\over2}^-)$}&\multirow{3}{*}{$2.59^{+0.19}_{-0.15}$}
&\multirow{6}{*}{$50\pm9$}&$\Lambda_c({1\over2}^-)\to \Sigma_c\pi\to\Lambda_c\pi\pi$&$g^S=0.58^{+0.81}_{-0.53}$&$ 0.1^{+0.5}_{-0.1}$&\multirow{2}{*}{}
&\multirow{3}{*}{$\Lambda_c(2595)$}
\\ \cline{5-7}
&&&&$\Lambda_c({1\over2}^-)\to\Sigma_c^*\pi\to\Lambda_c\pi\pi$&$g^D=0.69^{+0.61}_{-0.69}$
&$\sim0$&\multirow{1}{*}{$0.1^{+0.5}_{-0.1}$}&
\\ \cline{5-7}
&&&&$\Lambda_c({1\over2}^-)\to \Lambda_c\rho\to\Lambda_c\pi\pi$&$g^S=1.23^{+0.94}_{-1.23}$&$0.6^{+1.4}_{-0.6}\times10^{-3}$&&
\\ \cline{2-3} \cline{5-9}
&\multirow{4}{*}{$\Lambda_c({3\over2}^-)$}&\multirow{4}{*}{$2.64^{+0.20}_{-0.14}$}&
&\multirow{2}{*}{$\Lambda_c({3\over2}^-)\to\Sigma_c^*\pi\to\Lambda_c\pi\pi$}&$g^S=0.52^{+0.71}_{-0.47}$
&$0.1^{+0.3}_{-0.1}$&
\\&&&&&$g^D=0.91^{+0.45}_{-0.51}$&$1.3^{+1.5}_{-1.3}\times10^{-3}$
&\multirow{1}{*}{}&\multirow{2}{*}{$\Lambda_c(2625)$}
\\ \cline{5-7}
&&&&$\Lambda_c({3\over2}^-)\to\Sigma_c\pi$&$g^D=1.58^{+0.76}_{-0.89}$&$1.3^{+7.}_{-1.3}\times10^{-3}$
&\multirow{1}{*}{$0.1^{+0.3}_{-0.1}$}&
\\ \cline{5-7}
    &&&&$\Lambda_c({3\over2}^-)\to \Lambda_c\rho\to\Lambda_c\pi \pi$ & $g^S=0.71^{+0.81}_{-0.44}$&$1.0^{+3.5}_{-1.0}\times10^{-3}$&&
\\ \cline{2-9}
&\multirow{3}{*}{$\Xi_c({1\over2}^-)$}&\multirow{3}{*}{$2.80^{+0.08}_{-0.08}$}
&\multirow{7}{*}{$40\pm10$}&$\Xi_c({1\over2}^-)\to \Xi_c^{\prime}\pi$&$g^S= 0.36^{+0.71}_{-0.36} $&$4.5^{+34.5}_{-~4.5}$&
&\multirow{3}{*}{$\Xi_c(2790)$}
\\ \cline{5-7}
&&&&$\Xi_c({1\over2}^-)\to\Xi_c^{*}\pi\to\Xi_c\pi\pi$&$g^D= 0.49^{+0.32}_{-0.49}$&$\sim0$&\multirow{1}{*}{$4.5^{+34.5}_{-~4.5}$}
\\ \cline{5-7}
&&&&$\Xi_c({1\over2}^-)\to\Xi_c\rho\to\Xi_c\pi\pi$&$g^S=1.16^{+0.73}_{-1.16}$&$2.7^{+4.4}_{-2.7}\times 10^{-3}$&
\\ \cline{2-3} \cline{5-9}
&\multirow{4}{*}{$\Xi_c({3\over2}^-)$}&\multirow{4}{*}{$2.84^{+0.08}_{-0.08}$}&
&$\Xi_c({3\over2}^-)\to\Xi_c^{\prime}\pi$&$g^D=1.58^{+0.39}_{-0.70}$&$ 5^{+4}_{-3}\times 10^{-2}$& &\multirow{4}{*}{$\Xi_c(2815)$}
\\ \cline{5-7}
&&&&\multirow{2}{*}{$\Xi_c({3\over2}^-)\to\Xi_c^{*}\pi$}&$g^S=0.32^{+0.74}_{-0.32}$&$ 2.2^{+21.9}_{-~2.2}$&\multirow{2}{*}{$2.2^{+21.9}_{-~2.2}$}&\multirow{3}{*}{}
\\&&&&&$g^D= 0.91^{+0.23}_{-0.30}$&$2.6^{+1.5}_{-2.6}\times 10^{-4}$&
\\ \cline{5-7}
&&&&$\Xi_c({3\over2}^-)\to\Xi_c \rho\to\Xi_c \pi \pi$&$g^S=0.53^{+0.85}_{-0.44}$&$0.8^{+4.6}_{-0.8}\times 10^{-3}$&
\\ \hline\hline
\multirow{12}{*}{$[\mathbf{\bar 3}_F,2,1,\rho]$}&\multirow{3}{*}{$\Lambda_c({3\over2}^-)$}&\multirow{3}{*}{$2.65^{+0.24}_{-0.18}$}
&\multirow{4}{*}{$90\pm9$}&$\Lambda_c({3\over2}^-)\to \Sigma_c\pi$&$g^D=1.78^{+0.58}_{-0.88}$&$0.02^{+4.18}_{-0.02}$& &\multirow{3}{*}{--}
\\ \cline{5-7}
&&&&\multirow{2}{*}{$\Lambda_c({3\over2}^-)\to\Sigma_c^*\pi\to\Lambda_c\pi\pi$}&$g^S=0.48^{+0.18}_{-0.44}$
&$0.6^{+28.4}_{-~0.6}$&\multirow{1}{*}{$0.6^{+49.0}_{-~0.6}$}
\\&&&&&$g^D=1.03^{+0.33}_{-0.54}$&$0.01^{+17.0}_{-0.01}$&&
\\ \cline{2-3} \cline{5-9}
&\multirow{2}{*}{$\Lambda_c({5\over2}^-)$}&\multirow{2}{*}{$2.74^{+0.20}_{-0.16}$}&
&$\Lambda_c({5\over2}^-)\to\Sigma_c\pi$&$g^D=1.68^{+0.54}_{-0.88}$&$0.1^{+2.0}_{-0.1}$
&\multirow{2}{*}{$0.1^{+2.0}_{-0.1}$}&\multirow{2}{*}{--}
\\ \cline{5-7}
&&&&$\Lambda_c({5\over2}^-)\to\Sigma_c^*\pi$&$g^D=0.16^{+0.05}_{-0.08}$&$1.8^{+6.5}_{-1.8}\times 10^{-3} $&&
\\ \cline{2-9}
&\multirow{4}{*}{$\Xi_c({3\over2}^-)$}&\multirow{4}{*}{$2.85^{+0.15}_{-0.11}$}
&\multirow{7}{*}{$80\pm5$}&
\multirow{2}{*}{$\Xi_c({3\over2}^-)\to\Xi_c^{\star}\pi$}&$g^S=0.23^{+0.13}_{-0.19}$&$1.7^{+3.0}_{-1.7}$
& &\multirow{4}{*}{--}
\\&&&&&$g^D=1.41^{+0.36}_{-0.38}$&$0.7^{+29.0}_{-~0.3}\times10^{-2}$&\multirow{2}{*}{$4.7^{+18.0}_{-~4.7}$}
\\ \cline{5-7}
&&&&$\Xi_c({3\over2}^-)\to\Xi_c\pi$&$g^D=3.04^{+0.71}_{-0.60}$&$2.8^{+12.3}_{-~2.8}$
&
\\ \cline{5-7}
&&&&$\Xi_c({3\over2}^-)\to\Xi_c^{\prime}\pi$&$g^D=2.44^{+0.63}_{-0.76}$&$0.2^{+3.0}_{-0.2}$&
\\ \cline{2-3}\cline{5-9}
&\multirow{3}{*}{$\Xi_c({5\over2}^-)$}&\multirow{3}{*}{$2.93^{+0.10}_{-0.09}$}&
&$\Xi_c({5\over2}^-)\to\Xi_c\pi$&$g^D=4.30^{+1.00}_{-0.82}$&$5.4^{+9.0}_{-5.4}$&\multirow{2}{*}{}&\multirow{3}{*}{--}
\\ \cline{5-7}
&&&&$\Xi_c({5\over2}^-)\to\Xi_c^{\prime}\pi$&$g^D=2.30^{+0.59}_{-0.71}$&$0.4^{+1.0}_{-0.4} $&\multirow{1}{*}{$5.8^{+10.0}_{-~5.8}$}
\\ \cline{5-7}
&&&&$\Xi_c({5\over2}^-)\to\Xi_c^{*}\pi$&$g^D= 0.22^{+0.06}_{-0.07}$&$0.9^{+4.0}_{-0.9}\times10^{-2}$&
\\ \hline\hline
\end{tabular}}
\label{tab:decayb3f}
\end{center}
\end{table*}

\section{Radiative decay properties}
\label{sec:3FD}

In this section we investigate the radiative decay properties of the $P$-wave charmed baryons belonging to the $SU(3)$ flavor $\mathbf{\bar 3}_F$ representation. The analysis focuses on the following radiative decay channels:
\begin{align}
\Lambda_c^{+}&\to\Lambda_c^0(\Sigma_c^{+},\Sigma_c^{*0})\gamma \, ,
\\
\Xi_c^{+}&\to\Xi_c^+(\Xi_c^{\prime+},\Xi_c^{*+})\gamma \, ,
\\
\Xi_c^{0}&\to\Xi_c^0(\Xi_c^{\prime0},\Xi_c^{*0})\gamma \, ,
\end{align}
whose transition amplitudes are~\cite{Ivanov:1999bk}:
\begin{align}\label{amp1}
&\mathcal{M}(X_c({1/2}^-)\to Y_c({1/2}^+)\gamma)
\\ \nonumber
&~~~~~~~~~~~~~~~~~
=\frac{1}{\sqrt 3}g\bar X_c[g^{\mu\nu} v \cdot q-v^\mu q^\nu]\gamma_\nu\gamma_5 Y_c\epsilon^*_\mu \, ,
\\
&\mathcal{M}(X_c({1/2}^-)\to Y_c({3/2}^+)\gamma)
\\ \nonumber
&~~~~~~~~~~~~~~~~~
=g\bar X_c[g^{\mu\nu} v \cdot q-v^\mu q^\nu]Y_c^{\nu}\epsilon^*_\mu \, ,
\\
&\mathcal{M}(X_c({3/2}^-)\to Y_c({1/2}^+)\gamma)
\\ \nonumber
&~~~~~~~~~~~~~~~~~
=g\bar X_c^{\nu}[g^{\mu\nu} v \cdot q-v^\mu q^\nu]Y_c\epsilon^*_\mu \, ,
\\
&\mathcal{M}(X_c({3/2}^-)\to Y_c({3/2}^+)\gamma)
\\ \nonumber
&~~~~~~~~~~~~~~~~~
=\frac{1}{\sqrt 3}g\bar X_c^\alpha[g^{\mu\nu} v \cdot q-v^\mu q^\nu]\gamma_\nu\gamma_5Y_{c\alpha}\epsilon^*_\mu \, ,
\\
&\mathcal{M}(X_c({5/2}^-)\to Y_c({3/2}^+)\gamma)
\\ \nonumber
&~~~~~~~~~~~~~~~~~
=g\bar X_c^{\alpha\nu}[g^{\mu\nu} v \cdot q-v^\mu q^\nu]Y_{c\alpha}\epsilon^*_\mu \, ,
\end{align}
where $\epsilon_\mu$ denotes the polarization vector associated with the photon field. The radiative decay width can be further calculated through
\begin{equation}
\Gamma(X_c\to Y_c\gamma)=\frac{1}{2J+1}\frac{|\vec q|}{8\pi M_{X_c}^2}\sum\limits_{spins} {|\mathcal{M}(X_c\to Y_c\gamma)|^2}\, ,
\end{equation}
where $M_{X_c}$ is the mass of the initial baryon, and $\vec q$ is the three-momentum of the photon in the rest frame of the initial baryon.

As an example, we calculate the radiative decay of $\Xi_c^+({1/2}^-)$ belonging to the $[\mathbf{\bar 3}_F, 1, 1, \rho]$ doublet into $\Xi_c^{*+}(3/2^+)$ and $\gamma$. The four multiplets, $[\mathbf{\bar 3}_F, 0, 1, \rho]$, $[\mathbf{\bar 3}_F, 1, 1, \rho]$, $[\mathbf{\bar 3}_F, 2, 1, \rho]$, and $[\mathbf{\bar 3}_F, 1, 0, \lambda]$, can be analyzed in a similar manner. The obtained results for the $\Lambda_c$ and $\Xi_c$ baryons are summarized separately in Table~\ref{decayc3L} and Table~\ref{decayc3X}.

To investigate the above example, we consider the two-point correlation function:
\begin{align} \nonumber
\Pi^\mu(\omega, \, \omega^\prime) &=\int d^4 x e^{-i k \cdot x} \langle 0 | J_{\Xi_c^{+}[{\frac{1}{2}^-}],1,1,\rho}(0) \bar J^\mu_{\Xi_c^{*+}}(x) | \gamma \rangle
\\
&={1+v\!\!\!\slash\over2} \epsilon^*_\mu G_{\Xi_c^{+}[{1\over2}^-] \rightarrow \Xi_c^{*+}\gamma} (\omega, \omega^\prime) \, .
\end{align}

\begin{table*}[ht]
\begin{center}
\renewcommand{\arraystretch}{1.5}
\caption{Radiative decay properties of the $P$-wave charmed baryons $\Lambda_c$ belonging to the $SU(3)$ flavor $\mathbf{\bar 3}_F$ representation.}
\setlength{\tabcolsep}{0.1mm}{
\begin{tabular}{c| c | c | c | c | c | c| c }
\hline\hline
 \multirow{2}{*}{Multiplets}&Baryon & ~~~~~~Mass~~~~~~ &~~~ Difference ~~~& \multirow{2}{*}{~~~Decay channels~~~}&Coupling Constants  & ~~Decay width~~&~~~~Total~~~~
\\ & ($j^P$) & ({GeV}) & ({MeV}) & &({$\rm GeV^{-1}$})&({keV})&({keV})
\\ \hline\hline
~~\multirow{2}{*}{$[\mathbf{\bar 3}_F, 0, 1, \rho]$}~~ &
~~\multirow{2}{*}{$\Lambda_c({1\over2}^-)$}~~     &\multirow{2}{*}{$2.61^{+0.13}_{-0.13}$}      &\multirow{2}{*}{-}
&$\Lambda_c^+({1\over2}^-)\to\Sigma_c^+\gamma$          &$1.40^{+0.24}_{-0.38}$  &$690^{+490}_{-460}$                     &\multirow{2}{*}{$1100^{+1000}_{-~840}$}
\\
&&&&$\Lambda_c^+({1\over2}^-)\to \Sigma_c^{*+}\gamma$   &$1.70^{+0.36}_{-0.55}$   &$440^{+590}_{-380}$                        &
\\ \hline
\multirow{6}{*}{$[\mathbf{\bar 3}_F, 1, 1, \rho]$} &
\multirow{3}{*}{$\Lambda_c({1\over2}^-)$}     &\multirow{3}{*}{$2.59^{+0.19}_{-0.15}$}      &\multirow{6}{*}{$50\pm 9$}
&$\Lambda_c^+({1\over2}^-)\to \Lambda_c^+\gamma$        &$0.059^{+0.020}_{-0.020}$   &$7.8^{+6.2}_{-1.4}$                &\multirow{3}{*}{$12^{+10.0}_{-~2.8}$}
\\
&&&&$\Lambda_c^+({1\over2}^-)\to\Sigma_c^+\gamma$         &$0.11^{+0.05}_{-0.03}$  & $3.0^{+3.4}_{-1.4}$               &
\\
&&&&$\Lambda_c^+({1\over2}^-)\to \Sigma_c^{*+}\gamma$  &$0.065^{+0.025}_{-0.019}$ &  $0.34^{+0.34}_{-0.17}$          &
\\ \cline{2-3} \cline{5-8}
&\multirow{3}{*}{$\Lambda_c({3\over2}^-)$}&\multirow{3}{*}{$2.64^{+0.20}_{-0.14}$} &
&$\Lambda_c^+({3\over2}^-)\to \Lambda_c^+ \gamma$         &$0.084^{+0.028}_{-0.024}$  &  $21^{+16}_{-11}$          &\multirow{3}{*}{$24^{+19}_{-13}$}
\\
&&&&$\Lambda_c^+({3\over2}^-)\to\Sigma_c^+\gamma$         &$0.079^{+0.027}_{-0.026}$  & $2.9^{+2.8}_{-1.4}$     &
\\
&&&&$\Lambda_c^+({3\over2}^-)\to \Sigma_c^{*+}\gamma$  &$0.094^{+0.032}_{-0.031}$  &   $0.56^{+0.57}_{-0.26}$       &
\\ \hline
\multirow{3}{*}{$[\mathbf{\bar 3}_F, 2, 1, \rho]$} &
\multirow{2}{*}{$\Lambda_c({3\over2}^-)$}     &\multirow{2}{*}{$2.65^{+0.24}_{-0.18}$}      &\multirow{3}{*}{$90\pm9$}
&$\Lambda_c^+({3\over2}^-)\to\Sigma_c^+\gamma$           &$0.86^{+0.94}_{-0.41}$   &$500^{+4200}_{-~500}$                    &\multirow{2}{*}{$520^{+5200}_{-~520}$}
\\
&&&&$\Lambda_c^+({3\over2}^-)\to \Sigma_c^{*+}\gamma$      &$0.36^{+0.53}_{-0.24}$   &$15^{+270}_{-15}$        &
\\ \cline{2-3} \cline{5-8}
&\multirow{1}{*}{$\Lambda_c({5\over2}^-)$}     &\multirow{1}{*}{$2.74^{+0.20}_{-0.16}$}      &
&$\Lambda_c^+({5\over2}^-)\to\Sigma_c^{*+}\gamma$           &$0.82^{+0.14}_{-0.20}$   &$130^{+1000}_{-~130}$                    &$130^{+1000}_{-~130}$
\\ \hline
\multirow{4}{*}{$[\mathbf{\bar 3}_F, 1, 0, \lambda]$} &
\multirow{2}{*}{$\Lambda_c({1\over2}^-)$}     &\multirow{2}{*}{$2.66^{+0.22}_{-0.19}$}      &\multirow{4}{*}{$30\pm 5$}
&$\Lambda_c^+({1\over2}^-)\to\Sigma_c^+\gamma$          &$1.90^{+0.17}_{-0.27}$   &$2800^{+2400}_{-1700}$        &\multirow{2}{*}{$3400^{+3300}_{-2200}$}
\\
&&&&$\Lambda_c^+({1\over2}^-)\to \Sigma_c^{*+}\gamma$      &$1.10^{+0.10}_{-0.14}$ &$640^{+860}_{-480}$          &
\\ \cline{2-3} \cline{5-8}
&\multirow{2}{*}{$\Lambda_c({3\over2}^-)$}&\multirow{2}{*}{$2.69^{+0.23}_{-0.19}$} &
&$\Lambda_c^+({3\over2}^-)\to\Sigma_c^+\gamma$          &$1.30^{+0.22}_{-0.10}$    &$1900^{+1400}_{-~950}$         &\multirow{2}{*}{$2000^{+1500}_{-1000}$}
\\
&&&&$\Lambda_c^+({3\over2}^-)\to \Sigma_c^{*+}\gamma$     &$0.53^{+0.06}_{-0.05}$   &$71^{+80}_{-47}$                          &
\\ \hline\hline
\end{tabular}}
\label{decayc3L}
\end{center}
\end{table*}

\begin{table*}[ht]
\begin{center}
\renewcommand{\arraystretch}{1.5}
\caption{Radiative decay properties of the $P$-wave charmed baryons $\Xi_c$ belonging to the $SU(3)$ flavor $\mathbf{\bar 3}_F$ representation.}
\setlength{\tabcolsep}{0.1mm}{
\begin{tabular}{c| c | c | c | c | c | c |c}
\hline\hline
\multirow{2}{*}{Doublets}&Baryon & ~~~~~~Mass~~~~~~ & ~~Difference~~ & \multirow{2}{*}{~~~Decay channels~~~} &Coupling Constants & ~~Decay width~~&~~~~Total~~~~
\\ & ($j^P$) & ({GeV}) & ({MeV}) & &({$\rm GeV^{-1}$})&({keV})&({keV})
\\ \hline\hline
~~\multirow{4}{*}{$[\mathbf{\bar 3}_F, 0, 1, \rho]$}~~ &
~~\multirow{4}{*}{$\Xi_c({1\over2}^-)$}~~     &\multirow{4}{*}{$2.81^{+0.06}_{-0.06}$}      &\multirow{4}{*}{-}
&$\Xi_c^+({1\over2}^-)\to\Xi_c^{\prime+}\gamma$          &$1.40^{+0.24}_{-0.38}$    &$2100^{+2800}_{-1300}$                         &\multirow{2}{*}{$4200^{+6000}_{-3100}$}
\\
&&&&$\Xi_c^+({1\over2}^-)\to \Xi_c^{*+}\gamma$     &$1.60^{+0.41}_{-0.39}$  &$2100^{+3200}_{-1800}$                         &
\\ \cline{5-8}
&&&&$\Xi_c^0({1\over2}^-)\to\Xi_c^{\prime0}\gamma$        &$0.11^{+0.02}_{-0.01}$   &$13.0^{+13.0}_{-~7.3}$                         &\multirow{2}{*}{$27^{+32}_{-17}$}
\\
&&&&$\Xi_c^0({1\over2}^-)\to \Xi_c^{*0}\gamma$     &$0.13^{+0.02}_{-0.01}$  &$14^{+19}_{-10}$                         &
\\ \hline
~~\multirow{12}{*}{$[\mathbf{\bar 3}_F, 1, 1, \rho]$} ~~&
~~\multirow{6}{*}{$\Xi_c({1\over2}^-)$}~~     &\multirow{6}{*}{$2.80^{+0.08}_{-0.08}$}      &\multirow{12}{*}{$40\pm 10$}
&$\Xi_c^+({1\over2}^-)\to \Xi_c^+\gamma$          &$0.081^{+0.038}_{-0.041}$  &$17^{+12}_{-~9}$     &\multirow{3}{*}{$39^{+30}_{-21}$}
\\
&&&&$\Xi_c^+({1\over2}^-)\to\Xi_c^{\prime+}\gamma$    &$0.14^{+0.04}_{-0.04}$   &$18^{+15}_{-10}$                         &
\\
&&&&$\Xi_c^+({1\over2}^-)\to \Xi_c^{*+}\gamma$    &$0.076^{+0.028}_{-0.024}$  &$3.3^{+3.1}_{-1.7}$                        &
\\  \cline{5-8}
&&&&$\Xi_c^0({1\over2}^-)\to \Xi_c^0\gamma$        &$0.15^{+0.06}_{-0.05}$  &$59^{+51}_{-35}$   &\multirow{3}{*}{$59^{+51}_{-35}$}
\\
&&&&$\Xi_c^0({1\over2}^-)\to\Xi_c^{\prime0}\gamma$     &$0.005^{+0.004}_{-0.005}$  &$0.02^{+0.07}_{-0.02}$            &
\\
&&&&$\Xi_c^0({1\over2}^-)\to \Xi_c^{*0}\gamma$    &$0.007^{+0.002}_{-0.001}$  &$0.03^{+0.02}_{-0.02}$           &
\\ \cline{2-3}\cline{5-8}
&~~\multirow{6}{*}{$\Xi_c({3\over2}^-)$}~~     &\multirow{6}{*}{$2.8^{+0.08}_{-0.08}$}
&&$\Xi_c^+({3\over2}^-)\to \Xi_c^+\gamma$         &$0.11^{+0.06}_{-0.06}$   &$39^{+61}_{-31}$     &\multirow{3}{*}{$52^{+73}_{-37}$}
\\
&&&&$\Xi_c^+({3\over2}^-)\to\Xi_c^{\prime+}\gamma$     &$0.093^{+0.037}_{-0.029}$   &$10^{+10}_{-~5}$                       &
\\
&&&&$\Xi_c^+({3\over2}^-)\to \Xi_c^{*+}\gamma$      &$0.11^{+0.03}_{-0.04}$ &$3^{+2}_{-2}$                      &
\\  \cline{5-8}
&&&&$\Xi_c^0({3\over2}^-)\to \Xi_c^0\gamma$      &$0.21^{+0.08}_{-0.07}$    &$140^{+130}_{-~74}$                &\multirow{3}{*}{$140^{+130}_{-~74}$}
\\
&&&&$\Xi_c^0({3\over2}^-)\to\Xi_c^{\prime0}\gamma$     &$0.009^{+0.001}_{-0.001}$   &$0.10^{+0.20}_{-0.10}$               &
\\
&&&&$\Xi_c^0({3\over2}^-)\to \Xi_c^{*0}\gamma$     &$0.010^{+0.003}_{-0.002}$  &$0.03^{+0.10}_{-0.03}$                     &
\\ \hline
~~\multirow{6}{*}{$[\mathbf{\bar 3}_F, 2, 1, \rho]$}~~
&~~\multirow{4}{*}{$\Xi_c({3\over2}^-)$}~~    &\multirow{4}{*}{$2.85^{+0.15}_{-0.11}$}      &\multirow{8}{*}{$80\pm5$}
&$\Xi_c^+({3\over2}^-)\to\Xi_c^{\prime+}\gamma$         &$0.81^{+0.16}_{-0.18}$  & $1100^{+2600}_{-~940}$                        &\multirow{2}{*}{$1200^{+2800}_{-1000}$}
\\
&&&&$\Xi_c^+({3\over2}^-)\to \Xi_c^{*+}\gamma$     &$0.33^{+0.08}_{-0.08}$  &  $45^{+170}_{-~44}$                       &
\\  \cline{5-8}
&&&&$\Xi_c^0({3\over2}^-)\to\Xi_c^{\prime0}\gamma$      &$0.062^{+0.003}_{-0.003}$  & $6.5^{+15.5}_{-~5.0}$                          &\multirow{2}{*}{$7.0^{+17.0}_{-~6.0}$}
\\
&&&&$\Xi_c^0({3\over2}^-)\to \Xi_c^{*0}\gamma$     &$0.028^{+0.001}_{-0.001}$  & $0.33^{+1.17}_{-0.33}$                       &
\\ \cline{2-3}\cline{5-8}
&~~\multirow{2}{*}{$\Xi_c({5\over2}^-)$}~~    &\multirow{2}{*}{$2.93^{+0.10}_{-0.09}$}     &
&$\Xi_c^+({5\over2}^-)\to\Xi_c^{*+}\gamma$        &$0.27^{+0.05}_{-0.07}$   & $200^{+290}_{-150}$                &$200^{+290}_{-150}$
\\ \cline{5-8}
&&&&$\Xi_c^0({5\over2}^-)\to \Xi_c^{*0}\gamma$     &$0.017^{+0.001}_{-0.001}$  &  $0.83^{+0.98}_{-0.54}$                       &$0.83^{+0.98}_{-0.54}$
\\  \hline
~~\multirow{8}{*}{$[\mathbf{\bar 3}_F, 1, 0, \lambda]$} ~~
&~~\multirow{4}{*}{$\Xi_c({1\over2}^-)$} ~~    &\multirow{4}{*}{$2.79^{+0.09}_{-0.10}$}      &\multirow{8}{*}{$30\pm 3$}
&$\Xi_c^+({1\over2}^-)\to\Xi_c^{\prime+}\gamma$     &$2.10^{+0.30}_{-0.55}$      &$3800^{+1500}_{-1900}$                          &\multirow{2}{*}{$4600^{+2000}_{-2400}$}
\\
&&&&$\Xi_c^+({1\over2}^-)\to \Xi_c^{*+}\gamma$    &$1.20^{+0.20}_{-0.28}$   & $820^{+470}_{-460}$                         &
\\  \cline{5-8}
&&&&$\Xi_c^0({1\over2}^-)\to\Xi_c^{\prime0}\gamma$     &$0.089^{+0.007}_{-0.008}$   & $6.8^{+15.3}_{-~6.8}$                       &\multirow{2}{*}{$11^{+22}_{-11}$}
\\
&&&&$\Xi_c^0({1\over2}^-)\to \Xi_c^{*0}\gamma$     &$0.051^{+0.004}_{-0.004}$  & $4.5^{+6.8}_{-4.1}$                       &
\\ \cline{2-3}\cline{5-8}
&~~\multirow{4}{*}{$\Xi_b({3\over2}^-)$}~~     &\multirow{4}{*}{$2.82^{+0.09}_{-0.10}$}
&&$\Xi_c^+({3\over2}^-)\to\Xi_c^{\prime+}\gamma$         &$1.50^{+0.20}_{-0.42}$  &    $2800^{+1000}_{-1500}$                        &\multirow{2}{*}{$2900^{+1000}_{-1500}$}
\\
&&&&$\Xi_c^+({3\over2}^-)\to \Xi_c^{*+}\gamma$      &$0.58^{+0.11}_{-0.12}$  &   $90^{+42}_{-38}$                       &
\\  \cline{5-7}
&&&&$\Xi_c^0({3\over2}^-)\to\Xi_c^{\prime0}\gamma$      &$0.063^{+0.005}_{-0.005}$   & $4.9^{+12.2}_{-~4.9}$                        &\multirow{2}{*}{$5.0^{+13.0}_{-5.0}$}
\\
&&&&$\Xi_c^0({3\over2}^-)\to \Xi_c^{*0}\gamma$     &$0.026^{+0.002}_{-0.002}$   &$0.18^{+0.42}_{-0.18}$                       &
\\ \hline\hline
\end{tabular}}
\label{decayc3X}
\end{center}
\end{table*}

\begin{widetext}
At the hadronic level, the function $G_{\Xi_c^{+}[{1\over2}^-] \rightarrow \Xi_c^{*+}\gamma}(\omega, \omega^\prime)$ has the following pole term
\begin{align}
G_{\Xi_c^+[{1\over2}^-] \rightarrow \Xi_c^{*+}\gamma} (\omega, \omega^\prime)
= g_{\Xi_c^+[{1\over2}^-] \rightarrow \Xi_c^{*+}\gamma} \times { f_{\Xi_c^+[{1\over2}^-]} f_{\Xi_c^{*+}} \over (\bar \Lambda_{\Xi_c^+[{1\over2}^-]} - \omega^\prime) (\bar \Lambda_{\Xi_c^{*+}} - \omega)}+\cdot\cdot\cdot \label{G0C}\, .
\end{align}
At the quark-gluon level, we calculate the two-point correlation function $\Pi^\mu(\omega, \omega^\prime)$ by utilizing the method of operator product expansion, and extract $G_{\Xi_c^{+}[{1\over2}^-] \rightarrow \Xi_c^{*+}\gamma}(\omega, \omega^\prime)$ as
\begin{align}
\label{Gt1}
&G_{\Xi_c^{+}[{1\over2}^-] \rightarrow \Xi_c^{*+}\gamma} (\omega, \omega^\prime)
\\ \nonumber
&=\int_0^\infty  {dt} \int_0^1 {du{e^{i(1 - u)\omega 't}}{e^{iu\omega t}}}\times 8\Bigg(-\frac{{{e_s}{f_{3\gamma }}{\psi ^\alpha }({u})u v\cdot q}}{{48{\pi ^2}{t^2}}}+\frac{{{e_s}{\phi _\gamma }({u})\chi uv\cdot q}}{{72}}\left\langle {\bar qq}\right\rangle \left\langle {\bar ss} \right\rangle+\frac{{{e_s}{\phi_\gamma }({u})\chi t^2 u v\cdot q}}{{1152}}\left\langle {g_s\bar q\sigma Gs} \right\rangle\left\langle {\bar ss} \right\rangle
\\\nonumber
&+\frac{{{e_u}{f_{3\gamma }}{\psi ^\alpha }({u})uv\cdot q}}{{48{\pi ^2}{t^2}}}- \frac{{{e_u}{\phi_\gamma }({u})\chi {m_s}uv\cdot q}}{{24{\pi ^2}{t^2}}}\left\langle {\bar qq} \right\rangle-\frac{{{e_u}{\phi _\gamma }({u})\chi uv\cdot q}}{{72}}\left\langle {\bar qq} \right\rangle \left\langle {ss} \right\rangle
+\frac{{{e_u}{f_{3\gamma }}{\psi ^\alpha }({u}){m_s}t^2 u v\cdot q}}{{1152}}\left\langle {\bar ss} \right\rangle
\\\nonumber
&-\frac{{{e_u}{\phi _\gamma }({u})\chi t^2uv\cdot q}}{{1152}}\left\langle {\bar qq} \right\rangle \left\langle {g_s\bar s\sigma Gs} \right\rangle\Bigg)
-\int_0^\infty {dt} \int_0^1 {du\mathcal{D}{\underline\alpha}{e^{i\omega't({\alpha_2} + u{\alpha_3})}}{e^{i\omega t(1 - {\alpha_2} - u{\alpha_3})}}}\Bigg(-
\frac{{{f_{3\gamma }}\mathcal A(\underline\alpha)v\cdot q}}{{24{\pi ^2}{t^2}}}+ \frac{{{f_{3\gamma }}\mathcal V(\underline\alpha)v\cdot q}}{{24{\pi ^2}{t^2}}}
\\\nonumber
&- \frac{{{f_{3\gamma }}\mathcal A(\underline\alpha)wv\cdot q}}{{24{\pi^2}{t^2}}} - \frac{{{f_{3\gamma }}\mathcal V(\underline\alpha)wv\cdot q}}{{8{\pi^2}{t^2}}}
-\frac{{i{f_{3\gamma }}\mathcal A(\underline\alpha)(v\cdot q)^2}}{{24{\pi ^2}{t}}} - \frac{{i{f_{3\gamma }}\mathcal V(\underline\alpha)(v\cdot q)^2}}{{24{\pi ^2}{t}}} - \frac{{i{f_{3\gamma }}\mathcal A(\underline\alpha)w(v\cdot q)^2}}{{24{\pi ^2}{t}}} - \frac{{i{f_{3\gamma }}\mathcal V(\underline\alpha)w(v\cdot q)^2}}{{24{\pi ^2}{t}}}
\\\nonumber
&+ \frac{{i{f_{3\gamma }}\mathcal A(\underline\alpha){\alpha_2}(v\cdot q)^2}}{{24{\pi ^2}{t}}} + \frac{{i{f_{3\gamma }}\mathcal V(\underline\alpha){\alpha _2}(v\cdot q)^2}}{{24{\pi ^2}{t}}} + \frac{{i{f_{3\gamma }}\mathcal{A}(\underline\alpha)w{\alpha _2}(v\cdot q)^2}}{{24{\pi ^2}{t}}} + \frac{{i{f_{3\gamma }}\mathcal{V}(\underline\alpha)w{\alpha _2}(v\cdot q)^2}}{{24{\pi ^2}{t}}}\\ \nonumber
&+ \frac{{i{f_{3\gamma }}\mathcal A(\underline\alpha)w{\alpha _3}(v\cdot q)^2}}{{24{\pi ^2}{t}}} + \frac{{i{f_{3\gamma }}\mathcal V(\underline\alpha)w{\alpha _3}(v\cdot q)^2}}{{24{\pi ^2}{t}}} + \frac{{i{f_{3\gamma }}\mathcal A(\underline\alpha){w^2}{\alpha _3}(v\cdot q)^2}}{{24{\pi ^2}{t}}} + \frac{{i{f_{3\gamma }}\mathcal V(\underline\alpha){w^2}{\alpha _3}(v\cdot q)^2}}{{24{\pi ^2}{t}}}\Bigg)(e_s-e_u)\, .
\end{align}
After performing the double Borel transformation to respectively transform the two variables $\omega$ and $\omega^\prime$ to be $T_1$ and $T_2$, we obtain
\begin{align}
\label{Gt2}
& g_{\Xi_c^{+}[{1\over2}^-] \rightarrow \Xi_c^{*+}\gamma} \times{ f_{\Xi_c^+[{1\over2}^-]} f_{\Xi_c^{*+}} \over(\bar \Lambda_{\Xi_c^+[{1\over2}^-]}-\omega^\prime)(\bar\Lambda_{\Xi_c^{*+}}-\omega)}
\\ \nonumber
&= 8\times\Bigg(\frac{{{e_s}\chi }}{{72}}\left\langle {\bar qq} \right\rangle \left\langle {\bar ss} \right\rangle {(iT)^2}{f_1}(\frac{{{\omega _c}}}{T})\frac{\partial }{{\partial {u_0}}}{\phi _\gamma }({u_0}){u_0}+\frac{{{e_s}\chi }}{{1152}}\left\langle {g_s\bar q\sigma Gq} \right\rangle \left\langle {\bar ss} \right\rangle \frac{\partial }{{\partial {u_0}}}{\phi _\gamma }({u_0}){u_0}
+ \frac{{{e_u}{f_{3\gamma }}{m_s}}}{{1152}}\left\langle {\bar ss} \right\rangle \frac{\partial }{{\partial {u_0}}}{\psi ^\alpha }({u_0}){u_0}
\\\nonumber
&- \frac{{{e_u}\chi }}{{1152}}\left\langle {\bar qq} \right\rangle \left\langle {g_s\bar s\sigma Gs} \right\rangle \frac{\partial }{{\partial {u_0}}}{\phi _\gamma }({u_0}){u_0}-\frac{{{e_s}{f_{3\gamma }}}}{{48{\pi ^2}}}{(iT)^4}{f_3}(\frac{{{\omega _c}}}{T})\frac{\partial }{{\partial {u_0}}}{\psi ^\alpha }({u_0}){u_0}
+\frac{{{e_u}{f_{3\gamma }}}}{{48{\pi ^2}}}{(iT)^4}{f_3}(\frac{{{\omega _c}}}{T})\frac{\partial }{{\partial {u_0}}}{\psi ^\alpha }({u_0}){u_0}
\\\nonumber
&- \frac{{{e_u}\chi }}{{72}}\left\langle {\bar qq} \right\rangle \left\langle {\bar ss} \right\rangle {(iT)^2}{f_1}(\frac{{{\omega _c}}}{T})\frac{\partial }{{\partial {u_0}}}{\phi _\gamma }({u_0}){u_0}
-\frac{{{e_u}\chi {m_s}}}{{24{\pi ^2}}}\left\langle {\bar qq} \right\rangle {(iT)^4}{f_3}(\frac{{{\omega _c}}}{T})\frac{\partial }{{\partial {u_0}}}{\phi _\gamma }({u_0}){u_0}\Bigg)
\\\nonumber
&-\Bigg(-\frac{{{f_{3\gamma }}}}{{8{\pi ^2}{u_0}}}{(iT)^4}{f_3}(\frac{{{\omega _c}}}{T})\int_0^{\frac{1}{2}}{d\alpha_2} \int_{\frac{1}{2}-\alpha_2}^{1-\alpha_2}{d\alpha_3}\Big(\frac{1}{{3{\alpha_3}}}\frac{\partial }{{\partial {a_3}}}\mathcal A(\underline\alpha) - \frac{1}{{3{a_3}}}\frac{\partial }{{\partial {a_3}}}\mathcal V(\underline\alpha)
+ \frac{{{u_0}}}{{3{a_3}}}\frac{\partial }{{\partial {\alpha_3}}}\mathcal A(\underline\alpha) + \frac{{{u_0}}}{{{a_3}}}\frac{\partial }{{\partial {a_3}}}\mathcal V(\underline\alpha)\Big)
\\\nonumber
&+\frac{{{f_{3\gamma }}}}{{24{\pi ^2}u_0^2}}{(iT)^4}{f_3}(\frac{{{\omega _c}}}{T})\int_0^{\frac{1}{2}}{d\alpha_2} \int_{\frac{1}{2}-\alpha_2}^{1-\alpha_2}{d\alpha_3}\Big(
 - \frac{1}{{{\alpha_3}}}\frac{{{\partial ^2}}}{{\partial \alpha_3^2}}\mathcal A(\underline\alpha) - \frac{1}{{{\alpha_3}}}\frac{{{\partial ^2}}}{{\partial \alpha_3^2}}\mathcal V(\underline\alpha) - \frac{{{u_0}}}{{{\alpha_3}}}\frac{{{\partial ^2}}}{{\partial \alpha_3^2}}\mathcal A(\underline\alpha) - \frac{{i{u_0}}}{{{\alpha_3}}}\frac{{{\partial ^2}}}{{\partial \alpha_3^2}}\mathcal V(\underline\alpha)
\\\nonumber
&+ \frac{{{\alpha _2}}}{{{\alpha_3}}}\frac{{{\partial ^2}}}{{\partial \alpha_3^2}}\mathcal A(\underline\alpha) + \frac{{{\alpha _2}}}{{{\alpha_3}}}\frac{{{\partial ^2}}}{{\partial \alpha_3^2}}\mathcal V(\underline\alpha)
 + \frac{{{u_0}{\alpha _2}}}{{{\alpha_3}}}\frac{{{\partial ^2}}}{{\partial \alpha_3^2}}\mathcal A(\underline\alpha) + \frac{{{u_0}{\alpha _2}}}{{{\alpha_3}}}\frac{{{\partial ^2}}}{{\partial \alpha_3^2}}\mathcal V(\underline\alpha)
+ \frac{{{u_0}{\alpha _3}}}{{{\alpha_3}}}\frac{{{\partial ^2}}}{{\partial \alpha_3^2}}\mathcal A(\underline\alpha) + \frac{{{u_0}{\alpha _3}}}{{{\alpha_3}}}\frac{{{\partial ^2}}}{{\partial \alpha_3^2}}\mathcal V(\underline\alpha)
\\\nonumber
&+ \frac{{u_0^2{\alpha _3}}}{{{\alpha_3}}}\frac{{{\partial ^2}}}{{\partial \alpha_3^2}}\mathcal A(\underline\alpha) + \frac{{u_0^2{\alpha _3}}}{{{\alpha_3}}}\frac{{{\partial ^2}}}{{\partial \alpha_3^2}}\mathcal V(\underline\alpha)\Big)\Bigg)(e_s-e_u)\, .
\end{align}

\end{widetext}
In the above formalism, $e_{u/d/s}$ is the charge of the $up/down/strange$ quark, and the light-cone distribution amplitudes are taken from Refs.~\cite{Ball:1998je,Ball:2006wn,Ball:2004rg,Ball:1998kk,Ball:1998sk,Ball:1998ff,Ball:2007rt,Ball:2007zt,Ball:2002ps}. 

\begin{figure}[hbt]
\begin{center}
\scalebox{0.9}{\includegraphics{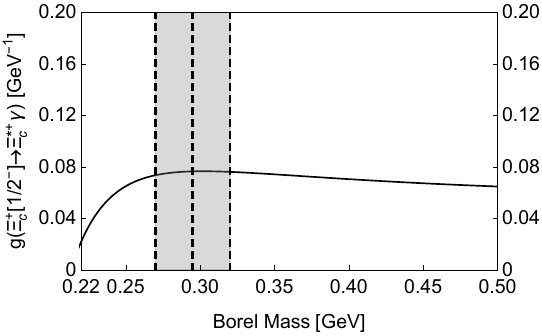}}
\caption{The coupling constant $g_{\Xi^+_c[\frac{1}{2}^-]\to\Xi_c^{*+}\gamma}$ as a function of the Borel mass $T$. Here the charmed baryon $\Xi_c^{+}({1/2}^-)$ belonging to the $[\mathbf{\bar 3}_F, 1, 1, \rho]$ doublet is investigated.}
\label{fig:es}
\end{center}
\end{figure}

The coupling constant $g_{\Xi_c^+[{1\over2}^-] \to \Xi_c^{*+}\gamma}$ depends on two free parameters, the threshold value $\omega_c$ and the Borel mass $T$. As listed in Table~\ref{tabmass}, we choose the Borel window $0.27$~GeV $<T<0.32$~GeV with $\omega_c = 1.79$ GeV. As depicted in Fig.~\ref{fig:es}, we obtain
\begin{align}
g_{\Xi_c^+[\frac{1}{2}^-]\to\Xi_c^{*+}[\frac{3}{2}^+]\gamma}&=0.076^{+0.028}_{-0.024} \, ,
\\
\Gamma_{\Xi_c^+[\frac{1}{2}^-]\to\Xi_c^{*+}[\frac{3}{2}^+]\gamma}&=3.3^{+3.1}_{-1.7}~\rm keV\, .
\end{align}

\section{Mixing between $[\mathbf{\bar3}_F, 1, 1, \rho]$ and $[\mathbf{\bar3}_F, 2, 1, \rho]$}
\label{sec:mixing}

HQET is an effective theory that works well for bottom baryons but is less suitable for charmed baryons. As a result, the three $J^P = 1/2^-$ charmed baryons can mix, and the three $J^P = 3/2^-$ charmed baryons can also mix. In particular, we investigate the mixing effect between the $[\mathbf{\bar3}_F, 1, 1, \rho]$ and $[\mathbf{\bar3}_F, 2, 1, \rho]$ doublets in this section. The procedure follows Ref.~\cite{Yang:2020zjl}, where the $P$-wave charmed baryons of the $SU(3)$ flavor $\mathbf{6}_F$ are studied.

We explicitly define the mixing states as
\begin{eqnarray}
\left(\begin{array}{c}
|\Lambda_c(3/2^-)\rangle_1\\
|\Lambda_c(3/2^-)\rangle_2
\end{array}\right)
&=&
\left(\begin{array}{cc}
\cos\theta & \sin\theta \\
-\sin\theta & \cos\theta
\end{array}\right)
\\ \nonumber && ~~~~~~~ \times
\left(\begin{array}{c}
|\Lambda_c(3/2^-),1,1,\rho\rangle\\
|\Lambda_c(3/2^-),2,1,\rho\rangle
\end{array}\right) \, ,
\\
\left(\begin{array}{c}
|\Xi_c(3/2^-)\rangle_1\\
|\Xi_c(3/2^-)\rangle_2
\end{array}\right)
&=&
\left(\begin{array}{cc}
\cos\theta & \sin\theta \\
-\sin\theta & \cos\theta
\end{array}\right)
\\ \nonumber && ~~~~~~~ \times
\left(\begin{array}{c}
|\Xi_c(3/2^-),1,1,\rho\rangle\\
|\Xi_c(3/2^-),2,1,\rho\rangle
\end{array}\right) \, ,
\end{eqnarray}
where $\theta$ is an overall mixing angle. We fine-tune it to be $\theta = 37\pm5^\circ$, which value is just the same as that used in Ref.~\cite{Yang:2021lce} for the mixing of the $[\mathbf{6}_F, 1, 1, \lambda]$ and $[\mathbf{6}_F, 2, 1, \lambda]$ doublets. The obtained results are summarized in Table~\ref{tab:result}. In particular, this mixing reduces the mass splitting within the $[\mathbf{\bar3}_F, 1, 1, \rho]$ doublet:
\begin{eqnarray*}
\\ \nonumber \Delta M_{[\Lambda_c, 1, 1, \rho]} &:& 50^{+9}_{-9}~{\rm MeV} \longrightarrow 36^{+17}_{-18}~{\rm MeV} \, ,
\\ \nonumber \Delta M_{[\Xi_c, 1, 1, \rho]} &:& 40^{+10}_{-10}~{\rm MeV} \longrightarrow 27^{+15}_{-15}~{\rm MeV} \, .
\end{eqnarray*}
These values show better agreement with the mass splitting between $\Lambda_c(2595)^+$ and $\Lambda_c(2625)^+$ as well as the mass splitting between $\Xi_c(2790)^{0/+}$ and $\Xi_c(2815)^{0/+}$~\cite{PDG2024}.

\begin{table*}[ht]
\begin{center}
\caption{Decay properties of the $P$-wave charmed baryons belonging to the $[\mathbf{\bar3}_F, 1, 1, \rho]$ and $[\mathbf{\bar3}_F, 2, 1, \rho]$ doublets. The first column lists charmed baryons categorized according to the heavy quark effective theory, and the third column lists the results after considering the mixing effect. Radiative decay widths are given in the sixth column for the excited $\Lambda_c^+$ and $\Xi_c^+$ baryons.}
\renewcommand{\arraystretch}{1.45}
\scalebox{0.90}{\begin{tabular}{   c|c | c | c | c | c | c | c}
\hline\hline
  \multirow{2}{*}{HQET state}&\multirow{2}{*}{Mixing}&\multirow{2}{*}{Mixed state} & Mass & Difference & ~~~~~~~~~~~\multirow{2}{*}{Decay channel}~~~~~~~~~~~ & Width  & \multirow{2}{*}{Candidate}
\\  &&&   ({GeV}) & ({MeV}) & & ({MeV}) &
\\ \hline\hline
$[\Lambda_c({1\over2}^-),1,1,\rho]$&--&$|\Lambda_c({1\over2}^-)\rangle$&$2.59^{+0.19}_{-0.15}$
&\multirow{4}{*}{$36^{+17}_{-18}$}&
$\begin{array}{l}
\Gamma^S\left(\Lambda_c({1/2}^-)\to\Sigma_c \pi\to\Lambda_c\pi\pi\right)=0.1^{+0.5}_{-0.1}~\rm MeV\\
\Gamma\left(\Lambda_c^+({1/2}^-)\to\Sigma_c^+ \gamma\right)=3.5^{+9.4}_{-3.5}~\rm keV\\
\Gamma\left(\Lambda_c^+({1/2}^-)\to\Sigma_c^{*+} \gamma\right)=0.48^{+3.04}_{-0.37}~\rm keV\\
\Gamma\left(\Lambda_c^+({1/2}^-)\to\Lambda_c^+ \gamma\right)=8.3^{+9.5}_{-6.6}~\rm keV
\end{array}$&$0.1^{+0.5}_{-0.1}$ &\multirow{1}{*}{$\Lambda_c(2595)$}
\\ \cline{1-4} \cline{6-8}
$[\Lambda_c({3\over2}^-),1,1,\rho]$&\multirow{6}{*}{$\theta={37\pm5^\circ}$}
&$|\Lambda_c({3\over2}^-)\rangle_1$
&$2.63^{+0.48}_{-0.47}$&&
$\begin{array}{l}
\Gamma^S({\Lambda_c({3/2}^-)\to\Sigma^{*}_c\pi\to\Lambda_c\pi\pi})=0.20^{+0.35}_{-0.13}~{\rm MeV} \\
\Gamma^D({\Lambda_c({3/2}^-)\to\Sigma^{*}_c\pi\to\Lambda_c\pi\pi})=3.2^{+2.2}_{-1.9}~{\rm keV} \\ \Gamma^D({\Lambda_c({3/2}^-)\to\Sigma_c\pi})=9.3^{+6.7}_{-5.6}~{\rm keV}\\
\Gamma({\Lambda^{+}_c({3/2}^-)\to\Sigma^{+}_c\gamma})=160^{+430}_{-100}~{\rm keV} \\
\Gamma({\Lambda^+_c({3/2}^-)\to\Sigma^{*+}_c\gamma})=5.9^{+20.1}_{-~4.6}~{\rm keV} \\
\end{array}$&$0.36^{+0.78}_{-0.23}$& \multirow{1}{*}{$\Lambda_c(2625)$}
\\ \cline{1-1} \cline{3-8}
$[\Lambda_c({3\over2}^-),2,1,\rho]$&&$|\Lambda_c({3\over2}^-)\rangle_2$&$2.67^{+0.46}_{-0.51}$
&\multirow{4}{*}{$80^{+18}_{-18}$}&
$\begin{array}{l}
\Gamma^S({\Lambda_c({3/2}^-)\to\Sigma^{*}_c\pi\to\Lambda_c\pi\pi})=0.1^{+2.1}_{-0.1}~{\rm MeV} \\
\Gamma^D({\Lambda_c({3/2}^-)\to\Sigma^{*}_c\pi\to\Lambda_c\pi\pi})=0.0^{+6.7}_{-0.0}~{\rm MeV} \\ \Gamma^D({\Lambda_c({3/2}^-)\to\Sigma_c\pi})=0.0^{+2.5}_{-0.0}~{\rm MeV}\\
\Gamma({\Lambda^{+}_c({3/2}^-)\to\Sigma^{+}_c\gamma})=0.3^{+6.7}_{-0.3}~{\rm MeV} \\
\Gamma({\Lambda^+_c({3/2}^-)\to\Sigma^{*+}_c\gamma})=0.0^{+0.4}_{-0.0}~{\rm MeV} \\
\end{array}$&$0.4^{+18.4}_{-~0.4}$&-
\\ \cline{1-4} \cline{6-8}
$[\Lambda_c({5\over2}^-),2,1,\rho]$&--&$|\Lambda_c({5\over2}^-)\rangle$&$2.74^{+0.20}_{-0.16}$&&
$\begin{array}{l}
\Gamma^D\left(\Lambda_c({5/2}^-)\to\Sigma_c^{*} \pi\right)=7^{+19}_{-~7}~\rm MeV\\
\Gamma^D\left(\Lambda_c({5/2}^-)\to\Sigma_c \pi\right)=2^{+12}_{-~2}~\rm MeV\\
\Gamma\left(\Lambda_c^+({5/2}^-)\to\Sigma_c^{*+} \gamma\right)=0.7^{+2.0}_{-0.7}~\rm MeV\\
\end{array}$&$10^{+33}_{-10}$&-
\\ \hline
$[\Xi_c({1\over2}^-),1,1,\rho]$&--&$|\Xi_c({1\over2}^-)\rangle$&$2.80^{+0.08}_{-0.08}$
&\multirow{5}{*}{$27^{+15}_{-15}$}&
$\begin{array}{l}
\Gamma^S({\Xi_c({1/2}^-)\to\Xi^{\prime}_c\pi})=4.5^{+34.5}_{-~4.5}~{\rm MeV} \\
\Gamma({\Xi^{+}_c({1/2}^-)\to\Xi^{+}_c\gamma})=18^{+25}_{-15}~{\rm keV} \\
\Gamma({\Xi^{+}_c({1/2}^-)\to\Xi^{*+}_c\gamma})=4.8^{+8.9}_{-4.8}~{\rm keV} \\
\end{array}$&$4.5^{+34.5}_{-~4.5}$&$\Xi_c(2790)$
\\ \cline{1-4}\cline{6-8}
$[\Xi_c({3\over2}^-),1,1,\rho]$&\multirow{6}{*}{$\theta={37\pm5^\circ}$}
&$|\Xi_c({3\over2}^-)\rangle_1$
&$2.83^{+0.22}_{-0.27}$&&
$\begin{array}{l}
\Gamma^S({\Xi_c({3/2}^-)\to\Xi^{*}_c\pi})=3.8^{+17.7}_{-~3.0}~{\rm MeV} \\
\Gamma^D({\Xi_c({3/2}^-)\to\Xi^{*}_c\pi})=1.8^{+0.2}_{-0.3}~{\rm keV} \\
\Gamma^D({\Xi_c({3/2}^-)\to\Xi^{\prime}_c\pi})=1.8^{+0.1}_{-0.0}~{\rm  keV} \\
\Gamma({\Xi^{+}_c({3/2}^-)\to\Xi^{\prime+}_c\gamma})=0.4^{+1.8}_{-0.4}~{\rm MeV} \\
\Gamma({\Xi^{+}_c({3/2}^-)\to\Xi^{*+}_c\gamma})=23^{+180}_{-~23}~{\rm keV} \\
\end{array}$&$4.2^{+19.5}_{-~3.4}$
& $\Xi_c(2815) $
\\ \cline{1-1}\cline{3-8}
$[\Xi_c({3\over2}^-),2,1,\rho]$&&$|\Xi_c({3\over2}^-)\rangle_2$&$2.86^{+0.33}_{-0.27}$
&\multirow{4}{*}{$70^{+8}_{-8}$}&
$\begin{array}{l}
\Gamma^S({\Xi_c({3/2}^-)\to\Xi^{*}_c\pi})=0.0^{+0.5}_{-0.0}~{\rm MeV} \\
\Gamma^D({\Xi_c({3/2}^-)\to\Xi^{*}_c\pi})=2^{+57}_{-~2}~{\rm keV} \\
\Gamma^D({\Xi_c({3/2}^-)\to\Xi^{\prime}_c\pi})=0.6^{+3.3}_{-0.6}~{\rm MeV} \\
\Gamma({\Xi^{+}_c({3/2}^-)\to\Xi^{\prime+}_c\gamma})=0.7^{+4.4}_{-0.7}~{\rm MeV} \\
\Gamma({\Xi^{+}_c({3/2}^-)\to\Xi^{*+}_c\gamma})=19^{+220}_{-~19}~{\rm keV} \\
\end{array}$&$1.4^{+8.2}_{-1.4}$&--
\\ \cline{1-4} \cline{6-8}
$[\Xi_c({5\over2}^-),2,1,\rho]$&--&$|\Xi_c({5\over2}^-)\rangle$
&$2.93^{+0.10}_{-0.09}$&&
$\begin{array}{l}
\Gamma^D({\Xi_c({5/2}^-)\to\Xi_c\pi})=7^{+3}_{-5}~{\rm MeV} \\
\Gamma^D({\Xi_c({5/2}^-)\to\Xi^{*}_c\pi})=0.01^{+0.06}_{-0.01}~{\rm MeV} \\
\Gamma^D({\Xi_c({5/2}^-)\to\Xi^{\prime}_c\pi})=0.5^{+0.5}_{-0.3}~{\rm MeV} \\
\Gamma({\Xi_c^+({5/2}^-)\to\Xi^{*+}_c\pi})=250^{+120}_{-120}~{\rm keV} \\
\end{array}$&$8.0^{+3.5}_{-5.3}$
&--
\\
\hline \hline
\end{tabular}}
\label{tab:result}
\end{center}
\end{table*}

\section{Summary and Discussions}
\label{secsummry}

In this paper we apply the light-cone sum rule method to systematically calculate the strong and radiative decay properties of the $P$-wave single charmed baryons belonging to the $SU(3)$ flavor $\mathbf{\bar3}_F$ representation within the framework of heavy quark effective theory. The obtained results are summarized in Tables~\ref{tab:decayb3f}, \ref{decayc3L}, and \ref{decayc3X} for the four $SU(3)$ flavor $\mathbf{\bar3}_F$ multiplets: $[\mathbf{\bar3}_F, 1, 0, \lambda]$, $[\mathbf{\bar3}_F, 0, 1, \rho]$, $[\mathbf{\bar3}_F, 1, 1, \rho]$, and $[\mathbf{\bar3}_F, 2, 1, \rho]$.

In particular, the total widths of the baryons belonging to the $[\mathbf{\bar3}_F, 1, 1, \rho]$ and $[\mathbf{\bar3}_F, 2, 1, \rho]$ doublets are relatively narrow, making these baryons accessible for experimental observation. Hence, we further consider the mixing effect between these two doublets, with the obtained results summarized in Table~\ref{tab:result}. The modified $[\mathbf{\bar3}_F, 1, 1, \rho]$ doublet consists of four excited charmed baryons: $\Lambda_c({1/2}^-)$, $|\Lambda_c(3/2^-)\rangle_1$, $\Xi_c({1/2}^-)$, and $|\Xi_c({3/2}^-)\rangle_1$. This doublet provides a consistent framework to explain the four excited charmed baryons $\Lambda_c(2595)^+$, $\Lambda_c(2625)^+$, $\Xi_c(2790)^{0/+}$, and $\Xi_c(2815)^{0/+}$ as a whole.

The modified $[\mathbf{\bar3}_F, 2, 1, \rho]$ doublet also consists of four excited charmed baryons: $|\Lambda_c(3/2^-)\rangle_2$, $\Lambda_c(5/2^-)$, $|\Xi_c(3/2^-)\rangle_2$, and $\Xi_c(5/2^-)$. Their masses, mass splittings within the same multiplets, and strong/radiative decay properties are summarized in Table~\ref{tab:result}, which may be investigated in future Belle II, BESIII, and LHCb experiments. As $\rho$-mode excitations, their experimental observation would not only validate our approach but also confirm the existence of the $\rho$-mode.

\section*{Acknowledgments}

This work is supported by
the National Natural Science Foundation of China under Grant No.~12075019,
the Jiangsu Provincial Double-Innovation Program under Grant No.~JSSCRC2021488,
and
the Fundamental Research Funds for the Central Universities.

\end{document}